\documentclass[twocolumn,english,english,floatfix,groupedaddress,superscriptaddress]{revtex4-1}
\usepackage[T1]{fontenc}
\usepackage[latin9]{inputenc}
\usepackage{color}
\pdfoutput=1
\usepackage{float}
\usepackage{booktabs}
\usepackage{amsmath}
\usepackage{amssymb}
\usepackage{graphicx}
\usepackage{esint}
\usepackage{color}
\RequirePackage[dvipsnames]{xcolor} 
\definecolor{RoyalBlue}{cmyk}{1, 0.80, 0, 0}
\usepackage{hyperref}
\hypersetup{
colorlinks=true, linktocpage=true, pdfstartpage=1, pdfstartview=FitV,%
breaklinks=true, pdfpagemode=UseNone, pageanchor=true, pdfpagemode=UseOutlines,%
plainpages=false, bookmarksnumbered, bookmarksopen=true, bookmarksopenlevel=1,%
hypertexnames=true, pdfhighlight=/O,hyperfootnotes=true,
urlcolor=RoyalBlue, 
citecolor=RoyalBlue, 
linkcolor=RoyalBlue, 
pagecolor=RoyalBlue,
pdftitle = {Exciton transport in thin-film cyanine dye J-aggregates},
pdfauthor = {$Stéphanie $Valleau}
}

\makeatletter

\newcommand{\lyxmathsym}[1]{\ifmmode\begingroup\def\b@ld{bold}
  \text{\ifx\math@version\b@ld\bfseries\fi#1}\endgroup\else#1\fi}

\providecommand{\tabularnewline}{\\}

 \@ifundefined{textcolor}{}
 {%
   \definecolor{BLACK}{gray}{0}
   \definecolor{WHITE}{gray}{1}
   \definecolor{RED}{rgb}{1,0,0}
   \definecolor{GREEN}{rgb}{0,1,0}
   \definecolor{BLUE}{rgb}{0,0,1}
   \definecolor{CYAN}{cmyk}{1,0,0,0}
   \definecolor{MAGENTA}{cmyk}{0,1,0,0}
   \definecolor{YELLOW}{cmyk}{0,0,1,0}
 }

\usepackage{amsthm}
\usepackage{braket}
\usepackage{graphics}
\usepackage{times}
\usepackage{bm}

\makeatother

\usepackage{babel}
\begin{document}

\title{Exciton transport in thin-film cyanine dye J-aggregates}

\author{Stéphanie Valleau}

\email{svalleau@fas.harvard.edu}

\author{Semion K. Saikin}

\author{Man-Hong Yung}

\author{Alán Aspuru Guzik}

\email{aspuru@chemistry.harvard.edu }

\address{Department of Chemistry and Chemical Biology, Harvard University,
Cambridge, Massachusetts 02138, USA}
\begin{abstract}
We present a theoretical model for the study of exciton dynamics in
J-aggregated monolayers of fluorescent dyes. The excitonic evolution
is described by a Monte-Carlo wave function approach which allows
for a unified description of the quantum (ballistic) and classical
(diffusive) propagation of an exciton on a lattice in different parameter
regimes. The transition between the ballistic and diffusive regime
is controlled by static and dynamic disorder. As an example, the model
is applied to three cyanine dye J-aggregates: TC, TDBC, and U3. Each
of the molecule-specific structure and excitation parameters are estimated
using time-dependent density functional theory. The exciton diffusion
coefficients are calculated and analyzed for different degrees of
film disorder and are correlated to the physical properties and the
structural arrangement of molecules in the aggregates. Further, exciton
transport is anisotropic and dependent on the initial exciton energy.
The upper-bound estimation of the exciton diffusion length in the
TDBC thin-film J-aggregate is of the order of hundreds of nanometers,
which is in good qualitative agreement with the diffusion length estimated
from experiments. 
\end{abstract}
\maketitle

\section{Introduction\label{sec:Introduction}}

In organic materials, excitons, quasiparticles of bound electron-hole
pairs, act as the intermediates between light (photons) and charge
(electrons and hole). Understanding which physical properties make
certain molecular aggregates optimal for exciton transfer is one of
the main current technological goals in organic material research.

In this article, we develop a computational model and employ it to
explore excitonic energy transport in a particular class of organic
materials: cyanine dye J-aggregates. 

Discovered over 50 years ago \cite{Jelley1936,Scheibe1936}, J-aggregates
are typically formed by organic fluorescent dye molecules and can
be identified spectroscopically by the narrowing and batochromic shift
(J-band) of the lowest electronic excitation relative to the monomer
band \cite{Wurthner2011,Higgins1996}. These structures are characterized
by the unique properties associated with their J-band: a large absorption
cross-section, short radiative lifetimes, a small Stokes shift of
the fluorescence line and efficient energy transfer within the aggregate
\cite{Wurthner2011}, which can be used for designing state-of-the-art
photonic devices. 

J-aggregates have been studied both experimentally \cite{Misawa1994,Moll1994,Mobius1988,Shirasaki2010,Tischler2009,Lochbrunner2011}
and theoretically \cite{Eisfeld2002,Michetti2008,Spano1989,Spano2010,Knoester1993}
and their applications range from being used as {}``reporter molecules''
in mitochondrial membrane potentials in living cells \cite{Smiley1991},
to photosensitizing silver halides in photography \cite{Jamesbook}.
Moreover, J-aggregates are imployed in dye-sensitized organic solar
cells which provide several advantages over inorganic solar cells
\cite{Saito1999,Sayama2002}. Recently, cyanine dye J-aggregates have
been combined with optical cavities \cite{Lidzey1999,Bellessa2004,Akselrod2010}
or coupled to quantum dots \cite{Walker2009,Walker2010} to form hybrid
systems. 

However, the current understanding of exciton transport properties,
even for the most ordered J-aggregates, is rather limited. This limitation
arises from the challenges encountered in the experimental characterization
of their structure \cite{Wurthner2011} and from the lack of information
on the dissipation processes. In an effort to overcome these difficulties,
the aim of our study is to provide a theoretical model, with a minimal
number of phenomenological parameters, which is useful for the determination
of the J-aggregate structure and able to describe the exciton dynamics
in hybrid excitonic-photonic and excitonic-electronic devices.

J-aggregates can be found in various structural arrangements including
one-dimensional, planar and cylindrical \cite{Wurthner2011} aggregates
each exhibiting different optical and exciton energy transfer properties
\cite{Eisfeld2007,Bakalis1997,Pugzlys2006}. The structure of liquid-crystal
cyanine dyes was initially studied using absorption and fluorescence
spectroscopy by Scheibe and Kandler \cite{Scheibe1938}, and more
recently using X-Ray diffraction and NMR by Harrison et al. \cite{Harrison1996,Harrison1996a}.
Nonetheless, the packing structure of J-aggregates remains unknown
\cite{Wurthner2011}. Different theoretical packing models for two-dimensional
arrays of these pseudo-cyanine dye (PIC) aggregates have been proposed
by Nakahara and Kuhn \cite{Nakahara1986}. In this study we focus
on modeling the exciton transport properties of two-dimensional (2D)
thin-films using cyanine dye J-aggregates such as those realized experimentally
in Ref. \cite{Bradley2005}. These highly efficient light absorbing
thin-films are employed in various opto-electronic systems for applications
such as lasers and optical switches \cite{Maltsev2000,Maltsev2002,McKeever2003}.
Exciton dynamics in these films, in general, possesses both ballistic
(quantum) and diffusive (classical) regimes and can be analyzed at
different levels of approximation. 

Experimentally, transport properties, such as diffusion coefficients
of excitons in organic materials can be obtained using indirect methods
only. These include exciton-exciton annihilation \cite{Shaw2010a,Akselrod2010,Lochbrunner2011},
photoluminescence quenching \cite{Lyons2005,Lunt2009}, transient
grating \cite{Salcedo1978} and photocurrent response \cite{Ghosh1978,BULOVI1995}.
In a recent study by Akselrod et. al. \cite{Akselrod2010} the singlet
exciton diffusion length in a 2D cyanine dye film has been estimated
to be of the order of 50 nm at room temperature which is more than
twice larger than the diffusion length measured in standard organic
semiconductor films \cite{Lunt2009}. 

Theoretically, exciton transport has been studied using a classical
hopping model \cite{Bulovic2006}. However, this approach is applicable
only in the weak F\"orster coupling regime \cite{Foersterbook},
where the exciton mobility is low. Beyond this regime, the tight-binding
Hamiltonian with classical noise model, proposed in the 70's by Haken,
Strobl, and Reineker \cite{Haken1972,Haken1973} allows for a unified
description of ballistic and diffusive exciton dynamics. For perfect
structures with translational symmetry this model can provide analytic
solutions for the moments of the exciton wave function \cite{Reineker1973}
that characterize exciton transport. Variations of this type of analysis
have been provided by others \cite{Ern1972,Reineker1981,Kenkre1981,Kuhne1981,Hoyer2010a}.
Incorporating a detailed computational ab-initio approach into this
model increases its accuracy by providing more insight on the specific
characteristics of exciton transport and by removing the limitation
on the structural symmetry. 

To demonstrate the applicablity of our model, we present a detailed
study of exciton diffusion in three types of cyanine-dye J-aggregates,
namely TC (5,5'-dichloro-3,3'-disulfopropyl thiacyanine), TDBC (5,6-dichloro-2{[}3-{[}5,6-dichloro-1-ethyl-3-(3-sulfopropyl)-2(3H)-benzimidazolidene{]}-1-propenyl{]}-1-ethyl-3-(3-sulfopropyl)
benzimidazolium hydroxide), and U3 \textcolor{black}{(3-{[}(2Z)-5-chloro-2-{[}((3E)-3-\{{[}5-chloro-3-(3-triethylammonium-sulfonatopropyl)-1,3-benzothiazol-3-ium-2-yl{]}methylene\}-2,5,5-trimethylcyclohex-1-en-1-yl))methylene{]}-1,3-benzothiazol-3(2H)-yl{]}
propane-1-sulfonate. The developed theoretical model provides a tool
for the analysis of transport properties and can be utilized in modeling
hybrid excitonic-photonic devices.}\textcolor{magenta}{{} }

Our findings indicate anisotropy in the exciton diffusion, the presence
of coherent dynamics at times shorter than tens of femtoseconds and
finally a dependence of transport on the specific molecular excitation
parameters. 

\begin{figure}
\begin{centering}
\includegraphics[width=0.6\columnwidth]{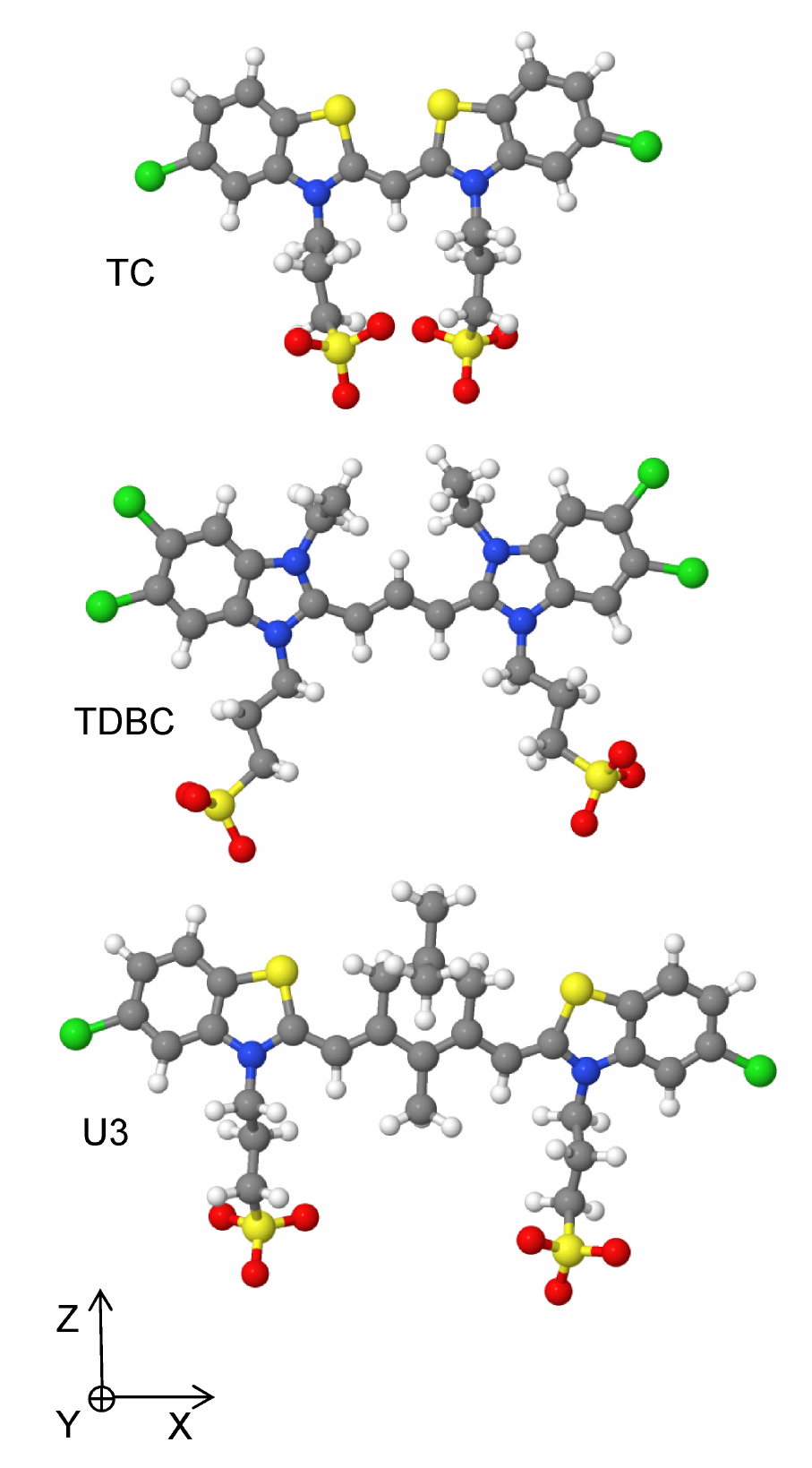}
\par\end{centering}

\caption{Structure of the monomer dye molecules TC, TDBC, and U3 which form
the aggregates. The full IUPAC names of the molecules are given in
the text.\label{fig:Structures of the cyanine dye monomers-1}}
\end{figure}

The paper is organized as follows. In Section \ref{sec:Model}, we
describe the theoretical model for the exciton dynamics. In particular,
we introduce the Hamiltonian of the system and the associated Langevin
equation. The static and dynamic noises, which represent the different
types of disorder present in J-aggregate films, are discussed. The
description of the model is completed with a derivation of the diffusion
equation. Section \ref{sec:Computational-details} includes details
of the calculation of the Hamiltonian's parameters and also gives
an overview of the Monte-Carlo Wave Function method (MCWF) employed
in the study of exciton propagation. In Section \ref{sec:Results}
we analyze exciton transport in thin-film J-aggregates of three different
cyanine dyes: TC, TDBC, and U3 (structures are shown in Fig. \ref{fig:Structures of the cyanine dye monomers-1}).
We conclude the study by summarizing our results in Section \ref{sec:conclusions}. 

\section{The Model }

\label{sec:Model}

\subsection{Hamiltonian and single exciton dynamics \label{sub:hamiltonian}}

We apply the general exciton theory developed previously for molecular
crystals and molecular aggregates \cite{Davydovbook,Agranovichbook}
to a specific system - a 2D monolayer J-aggregate of fluorescent dye
molecules. The exciton Hamiltonian of a 2D aggregate can be constructed
starting from that of single monomers by explicitly incorporating
intermolecular couplings. The multi-exciton space of the aggregate
consists of independent exciton manifolds that are defined by a specific
number of excitations in the system. The manifolds are coupled to
each other by exciton relaxation, annihilation or creation processes.
A more detailed discussion on the single-molecule Hamiltonian and
the exciton-manifolds is provided in Appendices \ref{sec: app_1}
and \ref{sec: app_2}. Here we constrain the exciton dynamics to the
single-exciton manifold and therefore assume that the exciton density
of the system is low. 
The Hamiltonian for a single exciton in a molecular aggregate can
be written as

\begin{equation}
\hat{H}=\hat{H}^{{\rm el}}+\hat{V}^{{\rm el-bath}}+\hat{H}^{{\rm bath}},\label{eq:Jagg Hamiltonian}
\end{equation}
where $\hat{H}^{{\rm el}}$ is the system Hamiltonian which includes
the electronic degrees of freedom, $\hat{V}^{{\rm el-bath}}$ is the
system-bath interaction Hamiltonian and $\hat{H}^{\mathrm{bath}}$
is the bath Hamiltonian. The electronic Hamiltonian in the site basis
can be expressed as

\begin{equation}
\hat{H}^{{\rm el}}=\sum_{n=1}^{N}\epsilon_{n}\ket{n}\bra{n}+\frac{1}{2}\sum_{n,m=1}^{N}J_{nm}\ket{n}\bra{m},\label{eq:Frenkel Exciton Hamiltonian}
\end{equation}
where $\epsilon_{n}$ are the energies of the electronic excitations
at each site $n$ and $J_{nm}$ are the couplings between electronic
transitions of monomers at sites $n$ and $m$, for an aggregate of
$N$ monomers. The energies $\epsilon_{n}$ are assumed to be
equal for all sites and disorder in these diagonal energies will be
included as described in Section \ref{sub:dephasing}. In Eq. \ref{eq:Frenkel Exciton Hamiltonian},
$\ket{n}=\ket{0...1_{n}...0}$ corresponds to the state where an exciton
is localized at the $n$-th molecule and all other molecules are in
their ground electronic state. In the aggregate, each monomer is modeled
as a two level system and the environment is assumed to be a harmonic
bath formed by the intra and intermolecular vibrations

\begin{equation}
\hat{H}^{{\rm bath}}=\sum_{n}\sum_{q}\omega_{q}\hat{b}_{qn}^{\dagger}\hat{b}_{qn},\label{eq:bath Hamiltonian}
\end{equation}
where $q$ runs over all vibrational modes. The system-bath interaction
term in the linear coupling limit is 

\begin{equation}
\hat{V}^{{\rm el-bath}}=\sum_{n}\ket{n}\bra{n}\sum_{q}\kappa_{q}\left(\hat{b}_{qn}^{\dagger}+\hat{b}_{qn}\right),\label{eq:coupling Hamiltonian}
\end{equation}
where $\hat{b}_{nq}^{\dagger}$ and $\hat{b}_{nq}$ are the (bosonic)
creation and annihilation operators for the bath modes at site $n$
and $\kappa_{q}$ is the coupling constant between the $q-th$ vibrational
mode and the electronic system, assumed to be equal for all sites. 

In general, there are two possible contributions to the coupling terms:
Dexter \cite{Dexter1953} and F\"orster interaction \cite{Foersterbook}.
If the wave-function overlap between interacting molecules is not
negligible, the Dexter interaction may give a significant contribution
to the $J_{nm}$ coupling term. While this term is dominant for triplet
exciton transport, singlet exciton transport can be accounted for
mostly by F\"orster coupling. Therefore, in this case, the major
contribution to the $J_{nm}$ coupling terms in Eq. \ref{eq:Frenkel Exciton Hamiltonian}
is due to the F\"orster interaction. For a pair of molecules $m$
and $n$, it is given by 
\begin{align}
J_{nm}^{{\rm F}} & =\frac{1}{4\pi\varepsilon_{0}\varepsilon}\int d\mathbf{r}_{1}\int d\mathbf{r}_{2}\braket{\phi_{ng}|\mathbf{r}_{1}}\braket{\phi_{me}|\mathbf{r}_{2}}\nonumber \\
 & \times\frac{1}{|\mathbf{r}_{1}-\mathbf{r}_{2}|}\braket{\mathbf{r}_{1}|\phi_{ne}}\braket{\mathbf{r}_{2}|\phi_{mg}},\label{eq:Foerster term}
\end{align}
where $\braket{\mathbf{r}|\phi_{ng}}$ and $\braket{\mathbf{r}|\phi_{ne}}$
are respectively the electron wave functions of the ground and excited
states of the $n$-th molecule in coordinate space, $\varepsilon_{0}$
is the vacuum permittivity and $\varepsilon$ is the permittivity
of the medium. In the following discussion, we will assume $\varepsilon=1$.
In practice, numerical calculations of the F\"orster term using Eq.
\ref{eq:Foerster term} can be computationally heavy, especially for
large structures. For J-aggregates, where the distance between stacked
molecules is comparable to the spatial extent of each molecule, it
is possible to use the extended dipole model \cite{Czikklely1970}.
Within this model, the F\"orster interaction between two particular
electronic transitions is parametrized by a transition charge, $q$,
and a transition dipole length $l$. The interaction term can therefore
be simplified and written as a sum of Coulomb interactions between
the transition charges located on different molecules,
\begin{equation}
J_{nm}^{{\rm F}}=\frac{q^{2}}{4\pi\varepsilon_{0}\varepsilon}\left(\frac{1}{r_{nm}^{++}}+\frac{1}{r_{nm}^{--}}-\frac{1}{r_{nm}^{+-}}-\frac{1}{r_{nm}^{-+}}\right)\label{eq:coupling term in extended dipole approximation-1}
\end{equation}
where $r_{nm}^{+-}$ is the distance between the charge $+q$ located
on the $n$-th molecule and the charge $-q$ located on the $m$-th
molecule. 

\begin{figure}
\begin{centering}
\includegraphics[width=0.95\columnwidth]{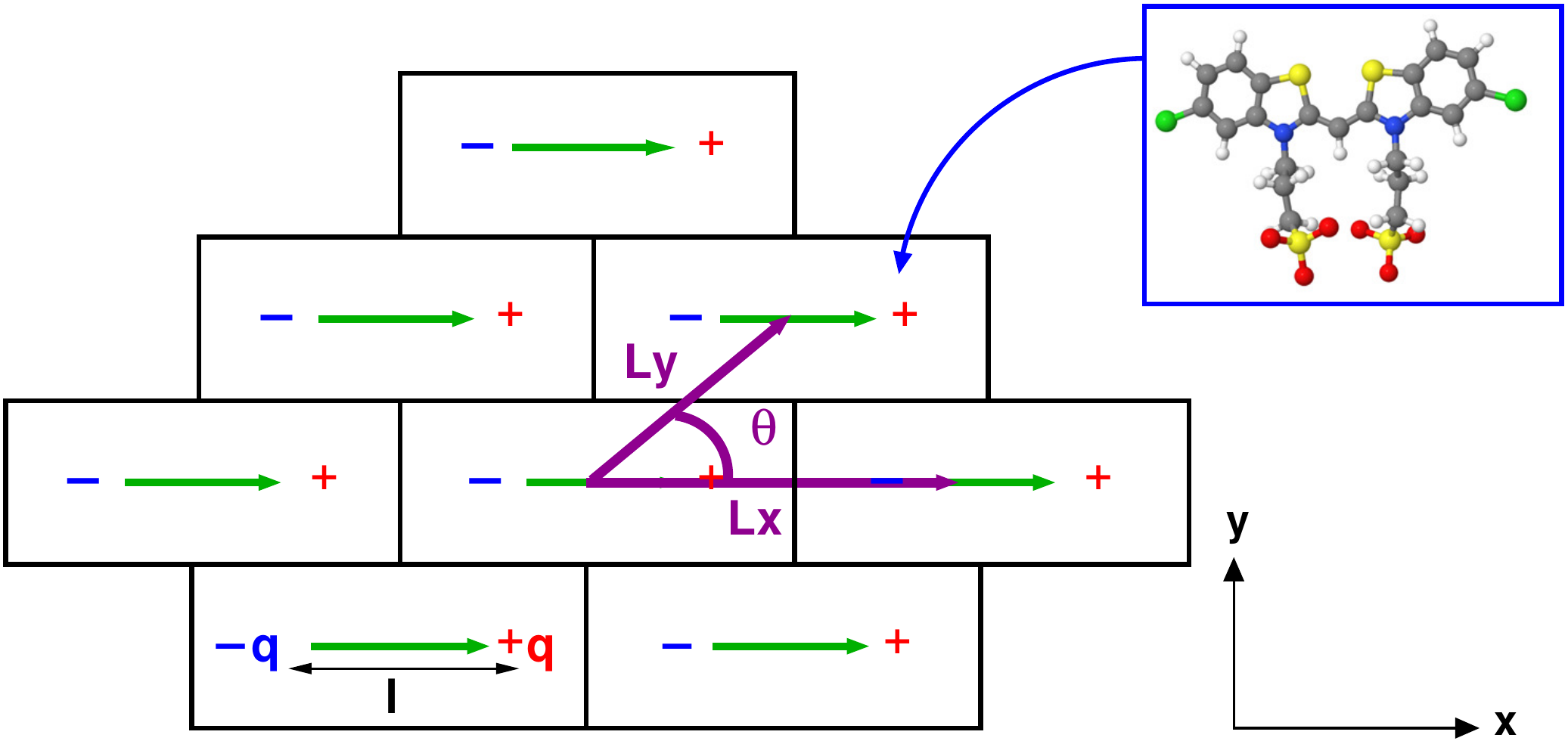}
\par\end{centering}

\caption{Brickstone lattice of a 2D J-aggregate. Each brick represents a molecule.
Green arrows indicate extended molecular transition dipoles and $l$
their length. The three lattice parameters are $\theta$, $L_{x}$
and $L_{y}$ which can be identified by the lattice vectors $\mathbf{L}_{x}$
and $\mathbf{L}_{y}$.\label{fig:Brickstone-lattice.-Green-1-1}}
\end{figure}
Fluorescent dyes may self-aggregate in a number of different structures
\cite{Jaggreview2011}. When dealing with two dimensional films of
cyanine dye aggregates, the brickstone model is one of most commonly
employed models which can account for the experimentally observed
optical properties of these aggregates. Therefore, in our work, the
molecular arrangement within the 2D layer is modeled as a brickstone
lattice \cite{Mobius1988}, as shown schematically in Fig. \ref{fig:Brickstone-lattice.-Green-1-1}.

In this model, the dye molecules are stacked parallel to each other
and subsequent rows are displaced by an angle $\theta$. The lattice
is characterized by two lattice vectors $\mathbf{L}_{x}$ and $\mathbf{L}_{y}$.
Given the Hamiltonian of the system, we proceed to derive an equation
for the quantum evolution of the system's wave function on this lattice. 

The quantum dynamics of excitons can be numerically simulated using
various methods, including density matrix evolution schemes \cite{Piilo2008,Rebentrost2009}
and also diffusion \cite{Roden2009,WFdiffusion1992} and quantum jump
\cite{Molmer1993,Ohta2006} wavefunction approaches. The latter are
based on propagating the system's wavefunction rather than the density
matrix and are therefore computationally more convenient to be employed
for modeling large systems. 

We define a quantum stochastic equation following the Langevin procedure,
as described in Ref. \cite{kampen2007}. Before including any source
of noise or relaxation, the equation of motion for a single exciton
in the aggregate is simply the Schr\"odinger equation

\begin{equation}
\frac{\partial\ket{\psi(t)}}{\partial t}=-\frac{i}{\hbar}\hat{H}^{{\rm el}}\ket{\psi(t)}\label{eq:schro_stat_dis}
\end{equation}
where $\ket{\psi(t)}$ is an exciton wave function in the single exciton
manifold and $\hat{H}^{{\rm el}}$ is defined in Eq. \ref{eq:Frenkel Exciton Hamiltonian}.
The initially excited exciton state will eventually decay back to
the ground state due to interaction with the environment. Processes
such as population relaxation and dephasing between excited states
and the ground state can be included as follows

\begin{equation}
\frac{\partial\ket{\psi(t)}}{\partial t}=\left(-\frac{i}{\hbar}\hat{H}^{{\rm eff}}+\sum_{m,\mu}\eta_{m}^{\mu}(t)\hat{C}_{m}^{\mu}\right)\ket{\psi(t)},\label{eq:full_langevin}
\end{equation}
where the effective Hamiltonian $\hat{H}^{{\rm eff}}=\hat{H}^{{\rm el}}-\frac{i\hbar}{2}\sum_{m,\mu}\left(\hat{C}_{m}^{\mu}\right)^{\dagger}\hat{C}_{m}^{\mu}$
includes a decay term and a stochastic fluctuation term $\eta_{m}^{\mu}(t)$
for site $m$ and channel $\mu$ which represents a dynamic noise
force and is introduced to conserve the norm of the wave function.
The $\hat{C}_{m}^{\mu}$ are Lindblad operators for two channels $\mu\in\{\wr,\phi\}$
(relaxation or dephasing) for each site $m$. In our case, these terms
can be expressed as 

\begin{align}
\hat{C}_{m}^{\wr} & =\sqrt{\Gamma^{\wr}}\ket{0}\bra{m}\nonumber \\
\hat{C}_{m}^{\phi} & =\sqrt{\Gamma^{\phi}}\left(\sum_{k\neq m}\ket{k}\bra{k}-\ket{m}\bra{m}\right).\label{eq:dephasing and relaxation}
\end{align}
where $\hat{C}_{m}^{\wr}$ describes single exciton relaxation and
$\hat{C}_{m}^{\phi}$ is introduced for dephasing processes, which
are associated with the relaxation $\Gamma^{\wr}$ and dephasing $\Gamma^{\phi}$
rates. The states $\ket{m}$ span over the single exciton manifold,
and the state $\ket{0}$ denotes the ground state, where no exciton
is present. The relaxation and dephasing rates are taken to be equal
for all sites, because the aggregate is constructed of identical molecules
which are assumed to be in identical local environments. It is easy
to see that with the noise operators written in the site basis this
model is equivalent to the Haken, Strobl and Reineker model \cite{Haken1972}.

\subsection{Static and dynamic noise\label{sub:dephasing}}

The environmental noise of the open quantum system needs to be introduced
into this J-aggregate model in its various manifestations. These include
fluctuations of site energies and of the site-to-site couplings as
well as a term which induces exciton relaxation. When studying transport
properties within the single exciton manifold, one is mostly interested
in the dynamics occurring on a timescale sufficiently shorter than
the exciton relaxation time, this corresponds to saying that over
the time scale of the dynamics, in Eq. 10, $\hat{C}_{m}^{\wr}\sim0$.
Thus, the exciton number can be assumed to be constant. Moreover,
for the sake of simplicity in the provided examples we do not consider
fluctuations of intermolecular interactions. Although these fluctuations
can give a noticeable contribution to the transport properties \cite{Ern1972}
the main effect of noise can be captured by the fluctuations of the
sole site energies.

Fluctuations of site energies can originate from low-frequency intramolecular
vibrations, molecular conformations, binding and un-binding events,
or charge fluctuations in a substrate or solvent which couples locally
to the electronic states of molecules. Such fluctuations have been
successfully modeled in natural molecular aggregates using molecular
dynamics methods \cite{Olbrich2011,Shim2012}. Detailed studies of
the physical origin of this noise is beyond the scope of the current
paper. Here, we only emphasize that one can introduce a distinction
between static and dynamic noise based on the correlation time characterizing
the fluctuations as compared to the exciton propagation time. For
instance if the conformation of a molecule changes on a nanosecond
timescale, the fluctuation of the site energy remains correlated during
the lifetime of an exciton which is of the order of tens of picoseconds.
Therefore, each exciton sees different, yet static local fluctuations.
In contrast, dynamic noise requires instantaneous fluctuations of
the site energy with a shorter correlation time, that is in a timescale
smaller or comparable to that of the energy transfer process.

Static noise can be accounted for in the Hamiltonian, Eq. \ref{eq:Frenkel Exciton Hamiltonian},
by introducing random shifts in the site energies. In our model, we
extract the random shifts from a Gaussian distribution 

\begin{equation}
f\left(\epsilon_{n}-\epsilon_{0}\right)=\frac{1}{\sqrt{2\pi\sigma^{2}}}e^{-(\epsilon_{n}-\epsilon_{0})^{2}/2\sigma^{2}}\label{eq:gaussian distribution-1}
\end{equation}
where $\sigma$ is the variance, and $\epsilon_{0}$ is the transition
energy of each isolated molecule. The static noise distribution is
assumed to be identical for all monomers. Other choices of static
disorder models are possible. For instance, a more general Levy distribution
can be used, which results in different state distributions and optical
properties \cite{Eisfeld2010}.

Dynamic noise directly enters the Schr\"odinger-Langevin equation,
Eq. \ref{eq:full_langevin} described in the previous section. The
requirement that the norm of $\ket{\psi(t)}$ is preserved imposes
certain properties on the dynamic noise force $\eta(t)$ \cite{kampen2007}.
First, the average of $\eta(t)$ should vanish, i.e., $\left\langle \eta(t)\right\rangle =0.$
Otherwise, one has to renormalize the unperturbed Hamiltonian in Eq.
\ref{eq:Frenkel Exciton Hamiltonian} to include the noise-induced
energy shifts. Second, the stochastic force $\eta(t)$, which is a
result of multiple uncorrelated microscopic movements of the molecules
constituting the aggregate lattice, does not have memory, i.e., $\left\langle \eta(t)\eta(t')\right\rangle \propto\Gamma^{\phi}\delta(t-t')$.
This corresponds to the Markov approximation, in the sense that the
system-bath interaction is assumed to be quasi instantaneous and successive
interaction events are not correlated. In general, the Markov approximation
holds, as long as the bath correlation time is much smaller than the
time over which one extracts properties of the system. Therefore it
is only necessary that the bath correlator be sharply peaked \cite{kampen2007,Louisell1990}.
Within this argumentation the dynamic noise can be characterized by
a single parameter $\Gamma^{\phi}$, which describes both the bath
dynamics and the system-bath coupling. While avoiding a detailed description
of the specific environment of a J-aggregate we can make the rather
qualitative assumption that $\Gamma^{\phi}$ is of the order of $k_{B}T$.
Such assumption is not strictly justified but can be intuitively explained
as following.

If we assume that the bath modes can be modeled as a set of harmonic
oscillators, we can define, within linear response theory, a system
bath interaction term linear in the displacement of the bath modes.
It follows that the bath correlator can be expressed as \cite{Breuerbook}

\begin{align}
\left\langle \eta(\tau)\eta(0)\right\rangle  & =\sum_{q}|\kappa_{q}|^{2}\langle\left(\hat{b}_{q}^{\dagger}(\tau)+\hat{b}_{q}(\tau)\right)\left(\hat{b}_{q}^{\dagger}(0)+\hat{b}_{q}(0)\right)\rangle\label{eq: bath correlator}\\
= & \int d\omega J\left(\omega\right)\left[\textrm{coth}\left(\frac{\hbar\omega\beta}{2}\right)\textrm{cos}\left(\omega\tau\right)-i\textrm{sin}\left(\omega\tau\right)\right]\nonumber 
\end{align}
where the bath operators are given in the interaction picture and
we have assumed that the noise correlator is the same for all sites
and that each bath is uncorrelated from that of other sites. The second
line corresponds to the approximation of a continuous bath spectrum
with a spectral density $J(\omega)$. Such approximation provides
a qualitative correspondence between the temperature and the dynamic
noise in the system. In general, a more quantitative analysis should
include explicitly the bath vibrational modes. By using the continuous
limit of the spectral density we intentionally simplified the model
making it independent of the specific molecular vibrational modes.
Several forms of bath spectral densities are used for modeling dissipative
quantum dynamics in molecular aggregates \cite{Renger2002,Malyshev2007,McKenzie2008,Rebentrost2009ENAQT}.
We choose to employ the Ohmic exponentially cutoff spectral density
$J(\omega)=\frac{\lambda}{\hbar\omega_{c}}\omega e^{-\frac{\omega}{\omega_{c}}}$
where $\lambda$ is the reorganization energy and $\omega_{c}$ is
the cutoff frequency. This spectral density can be integrated analytically
in Eq. \ref{eq: bath correlator} and it also has been used for the
simulation of exciton dynamics in natural molecular aggregates in
photosynthetic systems \cite{Cho2005,Rebentrost2009ENAQT}. In our
model, the correlator in Eq. \ref{eq: bath correlator} can be used
only in the limit when the time dependent characteristics of exciton
dynamics remain steady on timescales sufficiently longer than the
bath correlation time. In this case the temperature-dependent dephasing
rate can be defined as \cite{Rebentrost2009ENAQT,Adolphs2006} 
\begin{equation}
\Gamma^{\phi}=2\pi\frac{k_{B}T}{\hbar}\frac{\lambda}{\hbar\omega_{c}},\label{eq:dephasing rate}
\end{equation}
where physical properties of the bath and the system-bath coupling
are introduced through the slope of the spectral density at zero frequency
$\frac{dJ(\omega)}{d\omega}|_{\omega=0}=\frac{\lambda}{\hbar\omega_{c}}$
\cite{McKenzie2008}. Eq. \ref{eq:dephasing rate} is strictly valid
for $T\gg\frac{\hbar\omega_{c}}{k_{B}}$ \cite{Pachon2011}. If we
use, for example, values of $\omega_{c}=150\,\mathrm{cm^{-1}}$ and
$\lambda=35\,\mathrm{cm^{-1}}$, typical for the analysis of quantum
dynamics in photosynthetic systems \cite{Plenio2008,Rebentrost2009ENAQT},
the dephasing rate $\Gamma^{\phi}\approx1.4k_{B}T$. The reorganization
energy of J-aggregates is comparable to this value, for instance,
$\lambda_{\mathrm{TDBC}}=29\,\mathrm{cm^{-1}}$ \cite{Coles2010}.
By setting $\Gamma^{\phi}=k_{B}T$ we choose a lower bound for the
exciton dephasing rates. For room temperature we thus have $\Gamma^{\phi}\approx26\,\mathrm{meV}$.
In reality, there may be more sources of dissipation, and different
estimates for the bath spectral density \cite{Cho2005,McKenzie2008}
can give values of the dephasing rate, $\Gamma^{\phi}$ which are
several times larger than the value we use.

\subsection{Diffusion model\label{sub:diffusion_model}}

In this section, we consider methods for calculating the diffusion
constant $D$ from the transport properties of the excitonic system.
For classical Brownian motion, it is well-known that the diffusion
constant is related to the long-time limit of the second moment $\left\langle ({\bf r}(t)-{\bf r}_{0})^{2}\right\rangle $
evaluated for the initial condition where the particle is localized
at a single point ${\bf r}_{0}$ in space. Explicitly, it is given
by

\begin{equation}
D=\lim_{t\to\infty}\frac{1}{2td}\left\langle ({\bf r}(t)-{\bf r}_{0})^{2}\right\rangle ,\label{eq:diffusion constant}
\end{equation}
where $d$ is the dimension of the space. For excitonic systems, in
the first place, we will not assume that exciton motion can be described
as the motion of classical Brownian particles. Here our goal is to
show that the relationship above (Eq. \ref{eq:diffusion constant})
holds even for excitonic systems where a fully quantum mechanical
treatment is assumed. In other words, the non-equilibrium dynamics
of an initially localized exciton can indeed tell us about the diffusion
constant which describes the excitionic motion in (or near) equilibrium. 

This line of reasoning was first made by Scher and Lax \cite{Scher1973,Scher1973a}
for the calculation of the conductivity in doped semiconductors. It
was later applied to the calculation of the photoconductivity of organic
molecular crystals \cite{Kuhne1981,Reineker1981,Kenkre1981} and to
the bond-precolation model \cite{Odagaki1981}. The method developed
by Scher and Lax is based on linear response theory \cite{Kubo1957}.
To apply it to the excitonic system, we will have to imagine an external
perturbation which when applied, drives the motion of the excitons.
This can be physically achieved, e.g. through some non-resonant coupling
with an external field which creates a spatially inhomogeneous change
in the local potential term in the Hamiltonian (cf Eq. \ref{eq:Frenkel Exciton Hamiltonian}).
However, the actual implementation of this scheme is technically challenging
if not impossible. Furthermore, the formalism of Scher and Lax requires
the density matrix of the excitonic system to be an identity in thermal
equilibrium; as pointed out in Ref. \cite{Kenkre1981}, this may potentially
lead to internal inconsistency in the mathematical derivation. To
avoid these complications, in the following, we will present a simpler
way to obtain the relationship Eq. \ref{eq:diffusion constant} for
excitonic systems, without invoking linear response theory. This approach
is essentially the quantum extension of the classical approach described
in Ref. \cite{VanBeijeren1982}.

To get started, we consider a dilute system of excitons where their
reciprocal interaction can be ignored. We will trace the motion of
a {}``tagged'' exciton. The probability $P({\bf r},t)$ of finding
that exciton at location ${\bf r}$ and time $t$ is given by $P({\bf r},t)={\rm Tr[|{\bf r}\rangle\langle{\bf r}|}\hat{\rho}(t)]$,
where $\hat{\rho}(t)$ is the density matrix of the total system (i.e.
exciton plus the bath) at time $t$. Using the following identity

\begin{equation}
|{\bf r}\rangle\langle{\bf r}|=\frac{1}{(2\pi)^{d}}\int\limits _{-\infty}^{\infty}d{\bf k}e^{-i{\bf k}\cdot(\hat{{\bf r}}-{\bf r})},\label{eq:projector}
\end{equation}
where $d$ is the dimension, and $\mathbf{\hat{r}}\equiv\sum_{{\bf r'}}\mathbf{r'}|{\bf r}'\rangle\langle{\bf r'}|$
is the position operator for the exciton, we write the probability
of finding the exciton as 
\begin{equation}
P({\bf r},t)=\frac{1}{(2\pi)^{d}}\int\limits _{-\infty}^{\infty}d{\bf k}{\rm e^{i{\bf k}\cdot{\bf r}}}\widetilde{P}_{k}(t),\label{eq:probability}
\end{equation}
where $\widetilde{P}_{k}(t)\equiv{\rm Tr}[e^{-i{\bf k}\cdot\hat{{\bf r}}(t)}\hat{\rho}(0)]$,
$\hat{{\bf r}}(t)\equiv\hat{U}^{\dagger}(t)\hat{{\bf r}}\hat{U}(t)$,
where $\hat{U}(t)=e^{-i\hat{H}t}$ and $\hat{H}$ is defined in Eq.
\ref{eq:Jagg Hamiltonian}. 

Since we are considering the fluctuations of the exciton in some steady-state
(long-time $t\to\infty$) limit, measurable physical quantities, including
the diffusion constant, should not depend on the initial condition.
We can therefore choose the following initial state $\hat{\rho}(0)=\hat{\mathbb{I}}_{S}\otimes\hat{\rho}_{B}/{\rm Tr}(\hat{\mathbb{I}}_{S})$,
where $\hat{\mathbb{I}}_{S}$ is the identity matrix for the system,
i.e. the exciton, and $\hat{\rho}_{B}=e^{-\beta\hat{H}^{{\rm bath}}}/{\rm Tr}(e^{-\beta\hat{H}^{{\rm bath}}})$
is the density matrix of the bath, which is assumed to be in thermal
equilibrium. Now, inserting the resolution of the identity, $\hat{\mathbb{I}}_{S}=\sum_{{\bf r}_{0}}|{\bf r}_{0}\rangle\langle{\bf r}_{0}|$,
we can write $\widetilde{P}_{k}(t)=(1/{\rm tr}(\hat{\mathbb{I}}_{S}))\sum_{{\bf r_{0}}}e^{-i{\bf k\cdot{\bf r}_{0}}}{\rm Tr}[e^{-i{\bf k}\cdot\Delta\hat{{\bf r}}(t)}|{\bf r}_{0}\rangle\langle{\bf r}_{0}|\otimes\hat{\rho}_{B}]$
, where $\Delta\hat{{\bf r}}(t)\equiv\hat{{\bf r}}(t)-{\bf r}_{0}$.
Performing the cumulant expansion \cite{kampen2007} and keeping terms
up to the second-order, we obtain

\begin{equation}
\widetilde{P}_{k}(t)\approx\frac{1}{{\rm tr}(\hat{\mathbb{I}}_{S})}\sum_{{\bf r_{0}}}e^{-i{\bf k\cdot{\bf r}_{0}}}e^{-(1/2)\langle[{\bf k}\cdot\Delta\hat{{\bf r}}(t)]^{2}\rangle_{0}}\label{eq:probability_k}
\end{equation}
where $\langle[{\bf k}\cdot\Delta\hat{{\bf r}}(t)]^{2}\rangle_{0}\equiv{\bf {\rm Tr\{[}k}\cdot\Delta\hat{{\bf r}}(t)]^{2}|{\bf r}_{0}\rangle\langle{\bf r}_{0}|\otimes\rho_{B}\}$.
Then, we arrive at the diffusion equation

\begin{equation}
\frac{\partial}{\partial t}P({\bf r},t)\approx\sum_{i,j\epsilon\{x,y,z\}}D_{ij}(t)\frac{\partial^{2}}{\partial r_{i}\partial r_{j}}P({\bf r},t)\,,\label{eq:diffusion_like}
\end{equation}
where

\begin{equation}
D_{ij}(t)\equiv\frac{1}{2}\frac{d}{dt}\langle\Delta\hat{r}_{i}(t)\Delta\hat{r}_{j}(t)\rangle\equiv\frac{1}{2}\frac{d}{dt}M_{ij}^{(2)}(t)\label{eq:diffusion equation}
\end{equation}
is the time-dependent tensor of the diffusion coefficients. Here we
have defined the second moments $M_{ij}^{(2)}(t)\equiv\langle\Delta\hat{r}_{i}(t)\Delta\hat{r}_{j}(t)\rangle$.
In order for the diffusion coefficients $D_{ij}(t)$ to converge in
the long time limit ($t\to\infty$), the right-hand side of Eq. \ref{eq:diffusion equation}
should scale at most linearly in $t$. This result coincides with
that in Eq. \ref{eq:diffusion constant}, which is a special case
of isotropic diffusion.

\section{Computational details \label{sec:Computational-details}}

\subsection{Monomer properties}

Density-functional calculations of the molecular structure and electronic
excitation spectra were performed with the quantum-chemistry package
Turbomole, version 5.10. \cite{TMgeneral}. Triple-$\zeta$ valence-polarization
basis sets (def2-TZVP \cite{BSdef2}) were used together with the
hybrid functional of Perdew, Burke, and Ernzerhof (PBE0) \cite{PBE0}.
Dielectric properties of the medium where accounted for by using COSMO
\cite{Klamt1993} as implemented in Turbomole. 

To calculate the extended dipole parameters of the fluorescent dyes,
the HOMO and LUMO orbitals of the molecules were computed on a homogeneous
spatial grid. The grid steps are $dx=0.5$ Å, $dz=0.5$ Å in the plane
of the molecular backbone, and $dy=0.25$ Å in the direction orthogonal
to the backbone. The F\"orster, Eq. \ref{eq:Foerster term} and the
Dexter interactions between pairs of molecules were calculated for
center-to-center displacements scanned over $x=[-60,60]$ Å and $y=[-10,10]$
Å with step size $\Delta x=\Delta y=0.5$ Å .

\subsection{Monte Carlo wave function propagation \label{sub:MCWFM}}

Eq. \ref{eq:full_langevin} is a Markovian stochastic open quantum
system equation, which is equivalent to a Lindblad master equation
for the density matrix \cite{kampen2007}. As such one can evolve
single stochastic eigenfunction trajectories instead of the full density
matrix, using the Markov Monte-Carlo wave function method \cite{Molmer1993}. 

\begin{figure}
\begin{centering}
\includegraphics[clip,width=1\columnwidth]{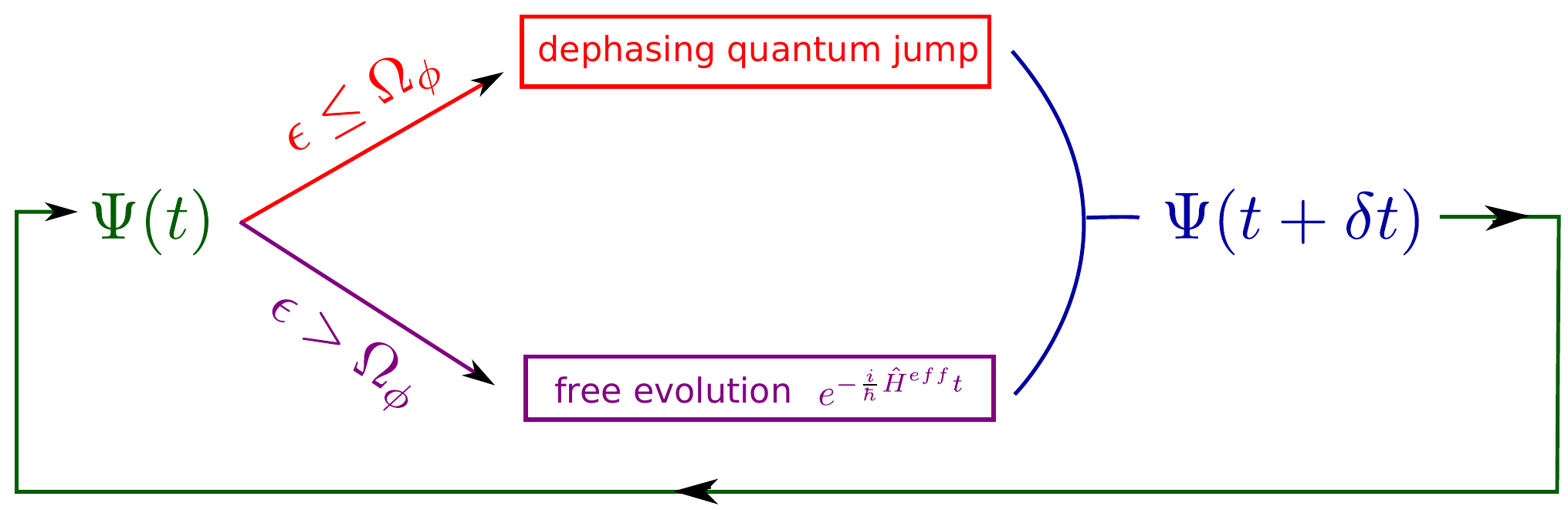}
\par\end{centering}

\caption{Monte Carlo Wavefunction stochastic dephasing jump evolution. Each
stochastic trajectory $\Psi(t)$ will evolve either by making a dephasing
quantum jump or according to a non-Hermitian effective hamiltonian
$\hat{H}^{\mathrm{eff}}$. The choice of making a dephasing jump is
determined by a Monte-Carlo type algorithm, where one extracts a random
number $\epsilon$ and compares it to the dephasing jump probability
$\Omega_{\phi}$. The procedure is repeated up to the final propagation
time. One then repeats the procedure to obtain an ensemble of trajectories
and average over this ensemble to obtain the desired observables.
\label{fig:Monte-Carlo-Wavefunction}}
\end{figure}

The initial wavefunction $\ket{\psi(0)}=\ket{00...1....0}$ is defined
as an exciton localized at the center of the lattice (i.e., in the
position $\left[\frac{n_{x}}{2};\frac{n_{y}}{2}\right]$) . At time
$t+\delta t$, $\ket{\psi(t+\delta t)}$ can be obtained from $\ket{\psi(t)}$
according to the stochastic scheme depicted in Fig. \ref{fig:Monte-Carlo-Wavefunction}.
 In particular, at each time step, if the stochastic variable extracted,
$\mathcal{\epsilon}$, is larger than the quantum dephasing jump probability
$\Omega_{\phi}$, the wavefunction will propagate freely under the
non-Hermitian effective Hamiltonian $\hat{H}^{\mathrm{eff}}$ introduced
in Eq. \ref{eq:full_langevin}

\begin{equation}
\ket{\psi(t+\delta t)}=\frac{1}{\sqrt{\mathcal{N}}}\left(1-\frac{i\hat{H}^{{\rm eff}}\delta t}{\hbar}\right)\ket{\psi(t)},\label{eq:psi_free_ev}
\end{equation}
where $\mathcal{N}=\sqrt{\braket{\psi(t+\delta t)|\psi(t+\delta t)}}$
is the normalization constant. However, if $\epsilon$ is smaller
than $\Omega_{\phi}$, a quantum dephasing jump will occur. The quantum
dephasing jump, a specific type of quantum jump, is described as a
flip of the sign of wavefunction's coefficient corresponding to a
site $m$ and corresponds to applying the dephasing jump operator
$\hat{C}_{m}^{\phi}$, Eq. \ref{eq:dephasing and relaxation}, to
the wavefunction \cite{Molmer1993}. The phase jump occurs in position
$N=\mathrm{round}(\frac{\epsilon}{\Omega_{\phi}}n_{x}n_{y})$ where
$\Omega_{\phi}$ is defined as $\Omega_{\phi}=\frac{\delta t}{2}\Gamma^{\phi}n_{x}n_{y}$,
and where $\Gamma^{\phi}$ was given in Eq. \ref{eq:dephasing rate}
while $\delta t$ is the time step which is assumed to be small enough
so that $\Omega_{\phi}\ll1.$ For the trajectories controlled by exciton
dephasing only, the effective Hamiltonian $\hat{H}^{{\rm eff}}$ is
the same as $\hat{H}_{0}$, thus the norm of the wavefuction in Eq.
\ref{eq:psi_free_ev} is conserved.

To analyze exciton transport properties of TC, TDBC and U3 aggregates
the exciton wavefunction was propagated on a lattice of 2601 monomers
($n_{x}=n_{y}=51$ ) for a total time of $t=100\,\mathrm{fs}$. The
quantum trajectories were averaged over 1000 different realizations
of static disorder (convergence on the populations was reached with
1000 realizations within an error $\leq4\%$). The time step in the
propagation was set to $\delta t=0.6\,\mathrm{as}$ thus the probability
of the quantum jump within the step is sufficiently smaller than 1,
and the wave function was collected at each femtosecond. The long
range interaction between the molecules has been accounted for within
the cutoff distance $l_{\mathrm{cutoff}}=6\cdot L_{x}$, more details
about this cutoff can be found in Section \ref{sub:parameters}. The
excitation was injected at the energy of the maximum of the J-band
and at time zero, only the central site (26,26) of the lattice was
excited. The results were obtained in general over the range $\Gamma^{\phi}=[20-110]\,\mathrm{meV}$
and $\sigma=[0-110]\,\mathrm{meV}$. Exciton diffusion coefficients
where estimated from a linear fit of the second moments of the exciton
distribution functions $M_{ii}^{(2)}$, as per Eq. \ref{eq:diffusion equation}.

\section{Results And discussion\label{sec:Results}}

\subsection{Model Parameters\label{sub:parameters}}

\subsubsection{Monomer calculations and absorption spectra}

\begin{figure}[h]
\centering{}\includegraphics[angle=-90,width=1\columnwidth]{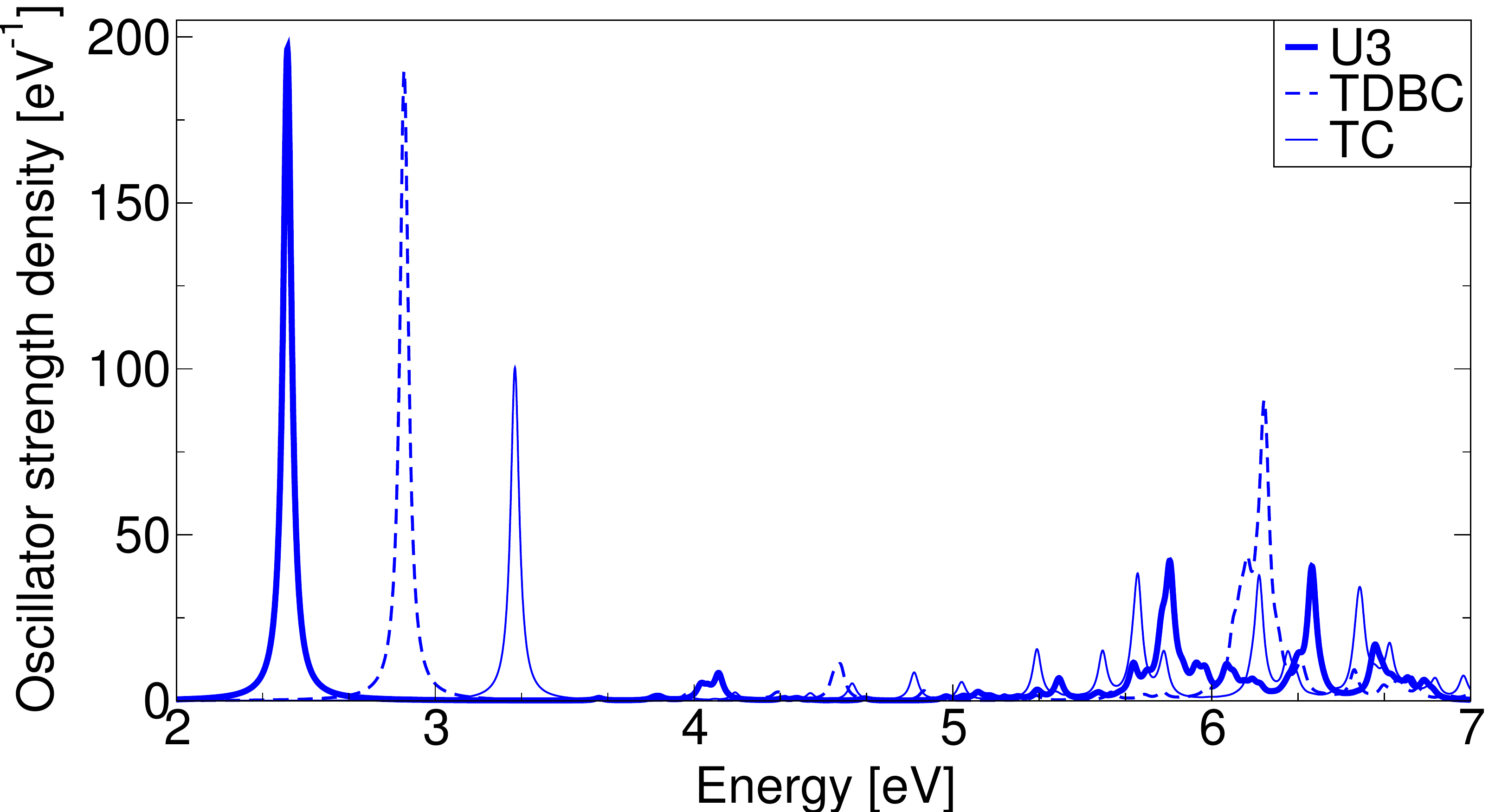}\caption{Computed spectra of electronic excitations in TC, TDBC, and U3 \emph{cis}-isomers.
The lines are broadened with Lorentzians of 20 meV linewidth. \label{fig:Absorption-spectra-of}}
\end{figure}

The structures of the TC, TDBC, and U3 cyanine dye molecules have
been optimized using density functional theory (DFT). In the computation,
we considered single-charged anions and assumed that the $\textrm{Na}^{+}$
ions were dissociated. For each molecule, although the conjugated
part of the structures remain almost planar, there are many conformations
which differ slightly by the orientations of the sulphonated group
side chains and are all closely spaced in energy. Examples of such
conformations for the \emph{cis}-isomers are shown in Fig. \ref{fig:Structures of the cyanine dye monomers-1}.
The difference between the ground-state energies of \emph{cis} and
\emph{trans}-isomers of the molecules is of the order of hundreds
of meV. We have chosen cis-isomers as our reference structure because
it is most likely that in this conformation, with the sulphonate groups
pointing towards the surface and binding chemically or physically
to it, that one would obtain the observed 2D monolayers \cite{Bradley2005}.
For each optimized molecular structure, we computed the 100 lowest
electronic excitations which fall in the energy range 2-7 eV. The
computed spectra of the molecules shown in Fig. \ref{fig:Structures of the cyanine dye monomers-1}
are provided in Fig. \ref{fig:Absorption-spectra-of}. The strong
lowest excitation can be accounted for by the HOMO to LUMO transition
for more than 98\%. Such transition is generally assigned to the lowest
absorption peak observed in the monomer spectra of fluorescent dyes.
The second electronic transition is separated from the lowest one
by about 0.7 eV, 1.0 eV, and 1.4 eV for TC, TDBC and U3 respectively.
Moreover, the oscillator strengths of these subsequent transitions
are about two orders of magnitude smaller than that of the lowest
one. These results support the two-level model provided that the static
energy disorder is of the order of 100 meV. The computed frequencies
of the lowest electronic transitions are systematically blue-shifted
by about several hundreds of meV as compared to the experimental values.
This shift typically occurs in DFT calculations with the PBE0 functional
\cite{DRapDFTbook2012}. Similarly to what was found for the ground
state energies, the lowest electron transition frequencies for \emph{trans}
and \emph{cis}-isomers differ by hundreds of meV's.

\subsubsection{Couplings}

\begin{figure}
\centering{}\includegraphics[width=1\columnwidth]{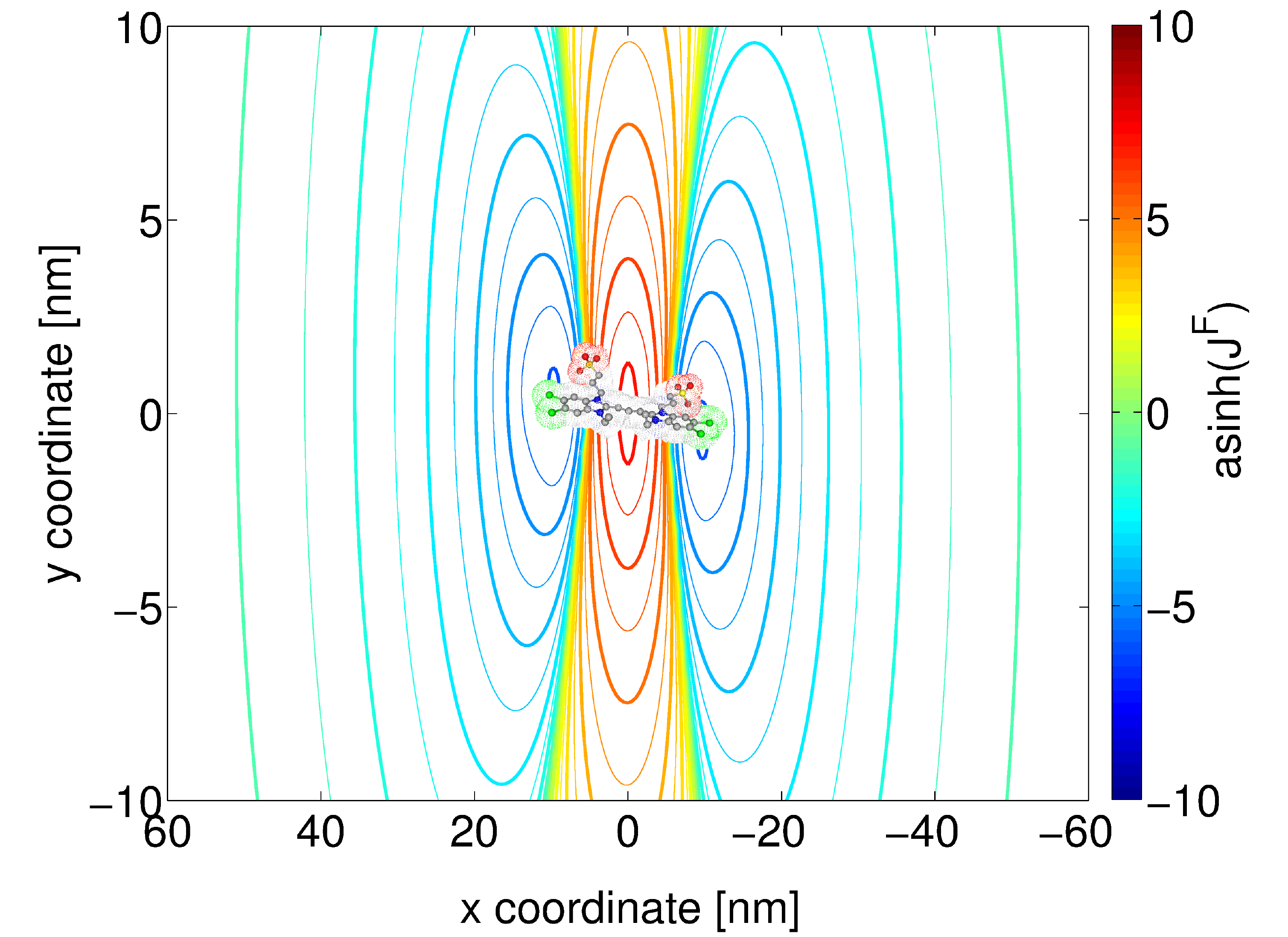}\caption{2D map of the F\"orster interaction between two TC molecules aligned
to each other. The axes show center-to-center displacement of the
molecules. The inset figure shows the orientation of the molecules.
The color map represents asinh$(J^{{\rm F}})$ to emphasize the long-range
behaviour of the interaction. The blue-green colors define the negative
frequency shift of the lowest electronic excitation, while the yellow-red
colors indicate the positive shift.\label{fig:Foerster-interaction}}
\end{figure}

The extended dipole parameters for the lowest electron excitation
are calculated within the Frontier Orbital Approximation (FAO) \cite{Fukui1952},
in which one assumes that only the HOMO-LUMO transition is involved.
To obtain the $l$ and $q$ parameters in the extended dipole coupling
formula Eq. \ref{eq:coupling term in extended dipole approximation-1},
we assume that the only type of interaction involved is F\"orster
interaction (Eq. \ref{eq:Foerster term}) and fit the interaction
between two molecules on the $x-y$ plane to the F\"orster results.
An example of the calculated interaction contour plot for a pair of
TDBC molecules is shown in Fig. \ref{fig:Foerster-interaction}. For
intermolecular distances larger than 2 Å, the profile reproduces the
interaction of two dipoles and can easily be fitted with the extended
dipole formula. The largest positive shift of the electron transition
is obtained when the molecules are displaced along the $y$-axis (direction
orthogonal to the backbone of the molecules). The largest negative
shift of the electron transition corresponds to the case when molecules
are displaced along the $x$-axis approximately by half of the length
of the molecule. These results are consistent with the extended dipole
model. 
 The computed properties of TC, TDBC, and U3 monomers are summarized
in Table \ref{tab:dye-properties}. We also computed Dexter couplings
and these were much smaller than the corresponding F\"orster terms
for distances of the order of the physical spacing between molecules
and were therfore neglected.

\begin{table}
\centering{}%
\begin{tabular}{cccccc}
\toprule 
Dye & $\Omega${[}eV{]} & $\mu${[}D{]} & $\mu_{\mathrm{HL}}$ {[}D{]} & $l$ {[}Å{]} & $q$ {[}\emph{e}{]}\tabularnewline
\midrule
TC & 3.3 & 8.9 & 8.3 & 9.1 & 0.20\tabularnewline
TDBC & 2.9 & 13.1 & 11.0 & 10.5 & 0.22\tabularnewline
U3 & 2.4 & 14.6 & 12.5 & 11.1 & 0.24\tabularnewline
\bottomrule
\end{tabular}\caption{Computed excitation properties of fluorescent dye molecules. $\Omega$
is the frequency of the electronic excitation, $\mu$ is the transition
dipole associated with the transition computed using TDDFT, $\mu_{\mathrm{HL}}$
is the transition dipole computed using HOMO-LUMO orbitals only, $l$
is the length of the extended dipole, and $q$ is the charge associated
with the extended dipole. \label{tab:dye-properties}}
\end{table}

\subsubsection{Lattice Parameters and absorption spectra}

The brickstone lattice parameters where determined as following. The
horizontal distance between monomers $L_{x}=\left|\mathbf{L}_{x}\right|$
was chosen to be the optimized length of the molecule in the \emph{cis}-geometry
$l_{0}^{cis}$ plus twice the Van der Waals radius of the chlorine
atom. In particular we chose the longitudinal Van der Waals radius
determined for a C-Cl type bond $r_{\mathrm{Cl}}=1.58\,\textrm{\AA}$,
as reported in Table 11 of \cite{Batsanov2001}. The angle between
monomers was set to $\theta={\rm tan^{-1}}\left(\frac{2L_{y}}{L_{x}}\right)$.
Having fixed $\theta$ and $L_{x}$ we then determined the vertical
distance between layers of monomers $L_{y}\mathrm{\cdot sin}\theta$
by fitting the theoretical position of the J-band in the absorption
spectra to the experimental result. All of these parameters are reported
in Table \ref{tab:Lattice_parameters} 
\begin{table}[H]
\begin{centering}
\begin{tabular}{ccccc}
\toprule 
Dye & $l_{0}^{cis}$$\textrm{\ensuremath{\left[\lyxmathsym{\AA}\right]}}$ & $L_{x}$$\textrm{\ensuremath{\left[\lyxmathsym{\AA}\right]}}$ & $L_{y}\mathrm{\cdot sin}\theta$$\left[\text{\AA}\right]$ & $\theta\left[\mathrm{rad}\right]$\tabularnewline
\midrule
TC & 15.01 & 18.17 & 3.815 & 0.421\tabularnewline
TDBC & 17.36 & 20.52 & 4.600 & 0.404\tabularnewline
U3 & 19.72 & 22.88 & 5.120 & 0.427\tabularnewline
\bottomrule
\end{tabular}
\par\end{centering}

\caption{Lattice parameters for the three molecules. $l_{0}^{cis}$ is obtained
from the DFT optimization of the molecular structure, $L_{y}$ parameters
were obtained from fitting the theoretical spectrum to the experimental
J-band shift. \label{tab:Lattice_parameters}}
\end{table}
The measured energies of the lowest electronic transitions for TC,
TDBC and U3 dyes in solution and also in the aggregated form are collected
in Table \ref{tab:Spectroscopic parameters}. 
\begin{table}[H]
\begin{centering}
\begin{tabular}{ccccc}
\toprule 
Dye & Monomer transition $\left[\mathrm{eV}\right]$ & J-band $\left[\text{eV}\right]$ & $\Delta$ $\left[\mathrm{eV}\right]$ & Ref.\tabularnewline
\midrule
TC & 2.900  & 2.613 & 0.287 & \textit{\emph{\cite{Kometani2001}}}\tabularnewline
TDBC & 2.396  & 2.115  & 0.281 & \cite{Akins2002} \tabularnewline
U3 & 1.864  & 1.571 & 0.293 & \textit{\emph{\cite{Wenus2007}}}\tabularnewline
\bottomrule
\end{tabular}
\par\end{centering}

\caption{Experimental data for electron excitations in monomer and J-aggregated
dyes as well as $\Delta$ the shift between monomer and J-band. \label{tab:Spectroscopic parameters}}
\end{table}

\begin{figure}[h]
\begin{centering}
\includegraphics[width=1\columnwidth]{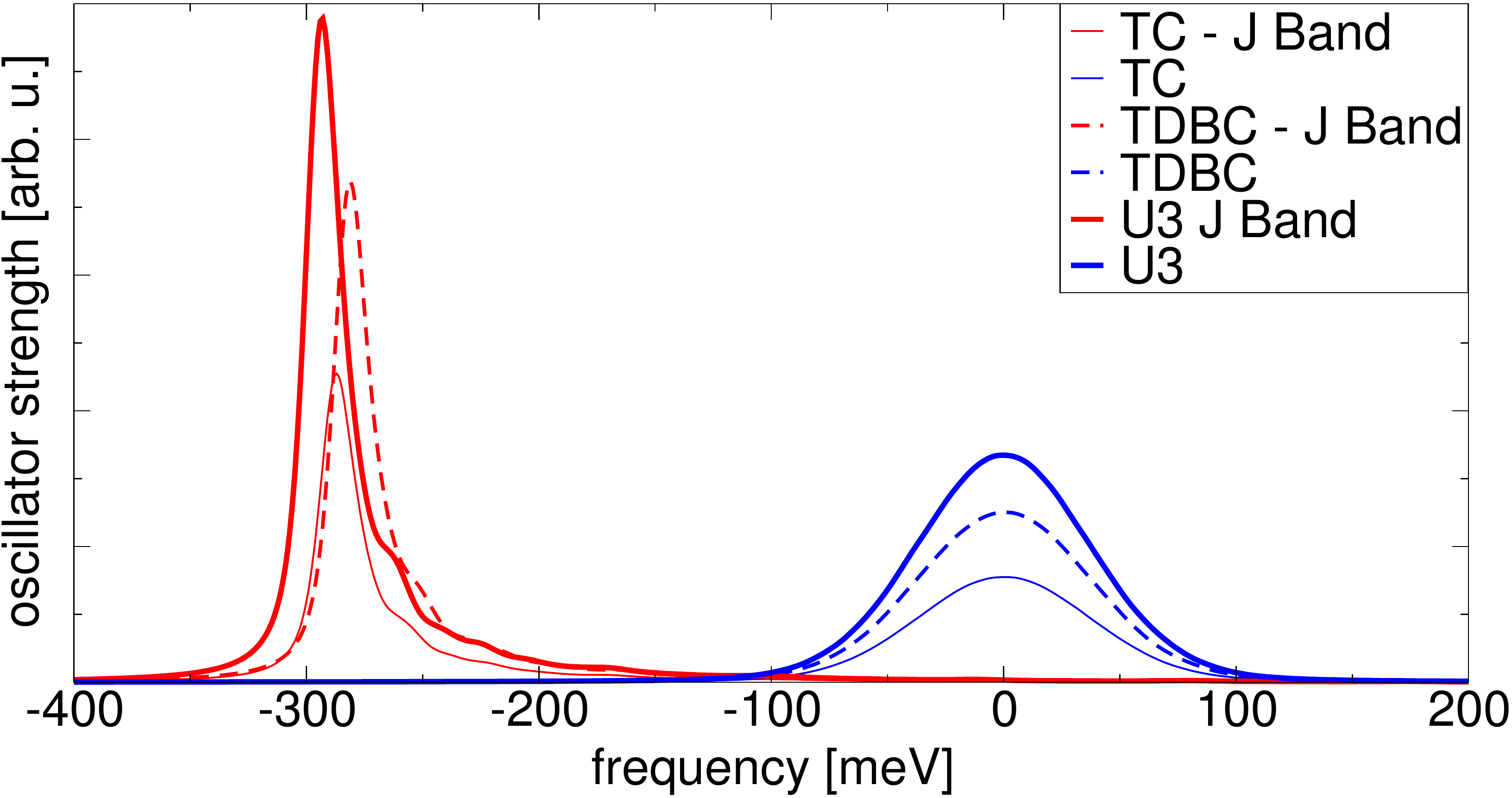}
\par\end{centering}

\caption{Calculated electronic excitation spectra of the three molecular aggregates,
TC, TDBC and U3, as compared to the electronic transitions of monomers
that form the aggregates. Zero frequency correspond to the energy
of a monomer excitation. The transition frequencies of the monomers
were taken from a Gaussian distribution of the width $70\,\mathrm{meV}$
to account for the static disorder. \label{fig:Calculated-absorption-spectrum}}
\end{figure}

To estimate the shifts of the J-band due to the molecular aggregation
the excitonic spectra of the aggregates have been computed by diagonalizing
the Hamiltonian, Eq. \ref{eq:Frenkel Exciton Hamiltonian}, for all
three dyes. The oscillator strength of a particular transition is
proportional to the square of the corresponding transition dipole.
To account for the static disorder the transition frequencies of the
monomers were taken from a Gaussian distribution of width $70\,\mathrm{meV}$.
In Fig. \ref{fig:Calculated-absorption-spectrum} the calculated spectra
of J-aggregates are shown as compared to the energies of single molecule
excitations. The peak positions where fitted to the experimental results.
To obtain the correct shift within 5\% of error, we determined that
the cut-off distance for the molecule-molecule interaction should
be not smaller than $l_{\mathrm{cutoff}}=6\cdot L_{x}$. We observe
the typical band narrowing of the J-band whereas the exact vibrational
structure of the monomer and the J-band cannot be captured with this
simple analysis and is beyond the scope of the present paper. Moreover,
the present analysis doesn't account for the line shift due to the
non-resonant (van der Waals) interactions. 

\subsection{Quantum exciton dynamics\label{sub: results qtm exciton dynamics}}

\begin{figure*}
\begin{centering}
\includegraphics[width=1.8\columnwidth]{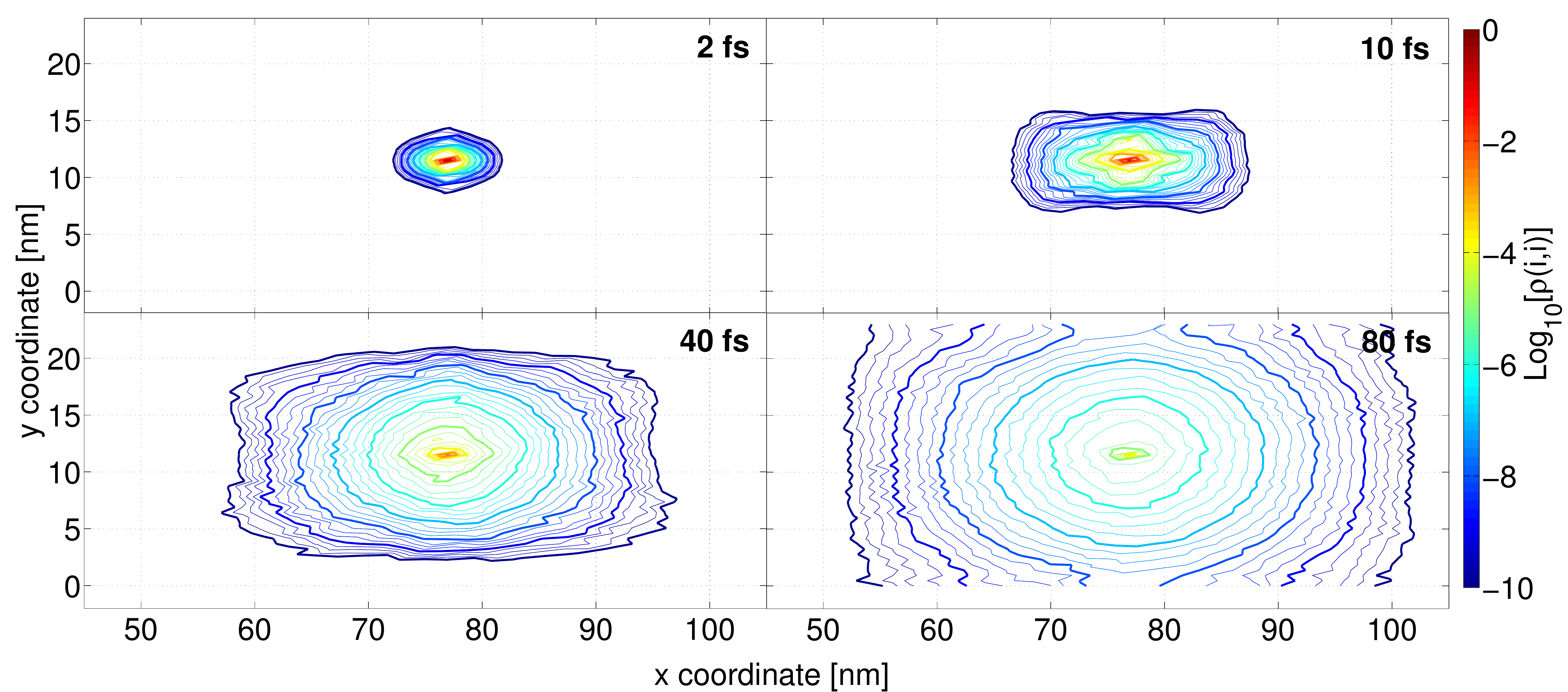}
\par\end{centering}

\caption{Contour plots of logarithm of exciton population in TDBC J-aggregate
projected onto coordinate space at 2, 10, 40 and $80\,\mathrm{fs}$.
Here, the static disorder is $\sigma=70\,\mathrm{meV}$ and the dynamic
disorder is $\Gamma^{\phi}=30\,\mathrm{meV}$. Population spreads
more rapidly in the $x$ direction than in the $y$ direction. This
behavior is observed for all studied values of dynamic and static
disorder and can be explained by the couplings between monomers as
described in the text. \label{fig:Pop1}}
\end{figure*}

\begin{figure*}
\begin{centering}
\includegraphics[width=1.8\columnwidth]{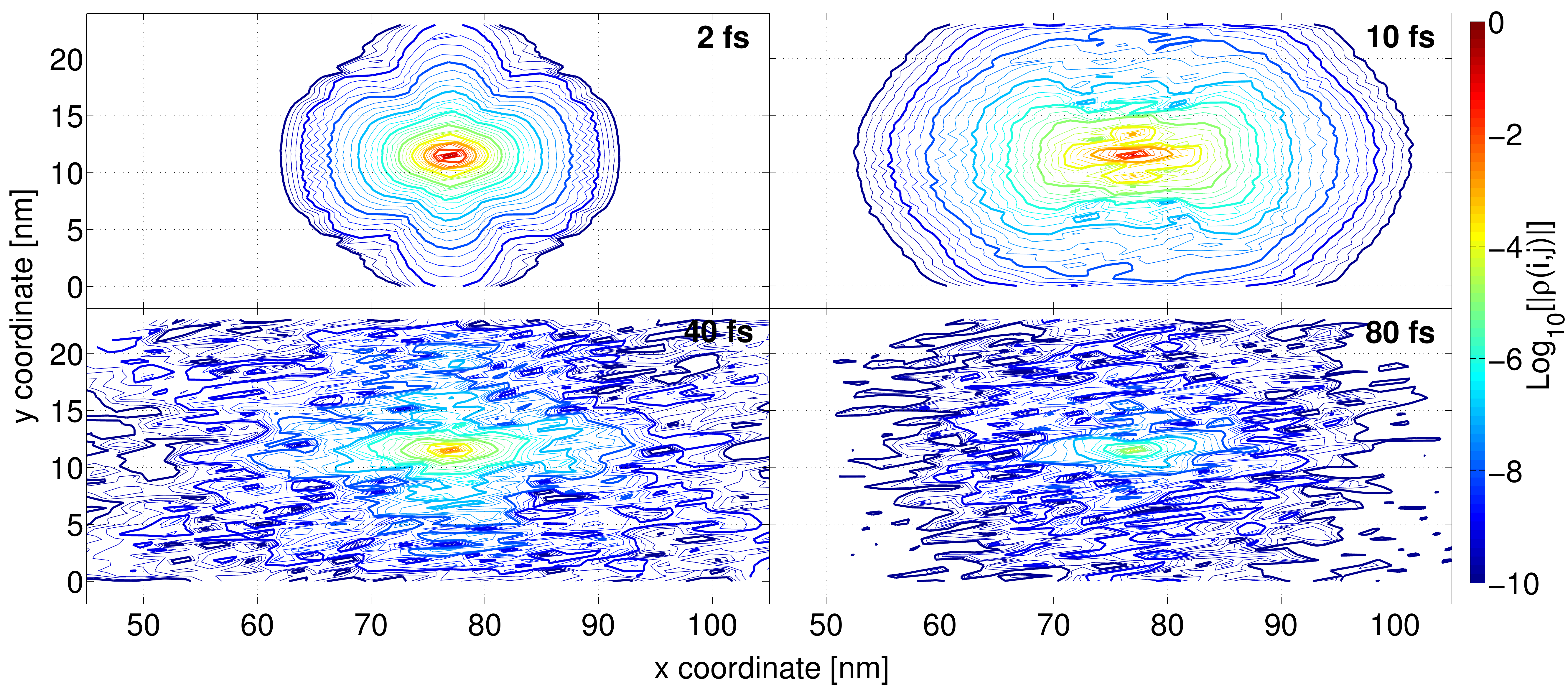}
\par\end{centering}

\caption{Contour plots of logarithm of exciton population in TDBC J-aggregate
projected onto coordinate space at 2, 10, 40 and $80\,\mathrm{fs}$.
Here, the static disorder is $\sigma=70\,\mathrm{meV}$ and the dynamic
disorder is $\Gamma^{\phi}=30\,\mathrm{meV}$. Coherences spread more
rapidly in the $x$ direction than in the $y$ direction. This behavior
is observed for all studied values of dynamic and static disorder
and can be explained by the couplings between monomers as described
in the text. Further, coherences spread and decay much more rapidly
than populations, as can be seen by comparing to Fig. \ref{fig:Pop1}.\label{fig:Cohe1}}
\end{figure*}

The developed model accounts for both coherent and incoherent properties
of exciton dynamics that can be monitored by following the spatial
distribution of the exciton population

\begin{equation}
P_{ij}(t)=\overline{\langle ij|\psi(t)\rangle\langle\psi(t)|ij\rangle},
\end{equation}
where $(i,j)$ is the pair of cartesian indices for a particular molecule
on the 2D lattice and the bar above the expression corresponds to
the ensemble average over quantum trajectories. In addition, to characterize
specifically the role of coherences in the exciton transport we analyze
the two-point one-time correlations between the central site (the
point of the exciton injection) and the current site $(i,j)$

\begin{equation}
C_{ij}(t)=\overline{|\langle ij|\psi(t)\rangle\langle\psi(t)|0\rangle|}.\label{eq:corr_nx_ny}
\end{equation}
A similar quantity has been used previously in Refs. \cite{Pullerits1997,Pullerits1999}
to estimate the exciton delocalization length in natural molecular
aggregates. If the exciton transport is completely incoherent and
represented by the hopping of exciton population between sites, the
correlation function, Eq. \ref{eq:corr_nx_ny}, should remain zero
for all times provided that no initial site-site correlations were
created. This is consistent with the conventional Bloch equations,
where the coherence dynamics and the population dynamics are separated.
In Figs. \ref{fig:Pop1} and \ref{fig:Cohe1} we show an example of
the population and coherence dynamics in TDBC J-aggregate when the
dynamic disorder is $\Gamma^{\phi}=30\,\mathrm{meV}$ and static disorder
is $\sigma=70\,\mathrm{meV}$. 

\begin{figure}[h]
\begin{centering}
\includegraphics[angle=-90,width=1\columnwidth]{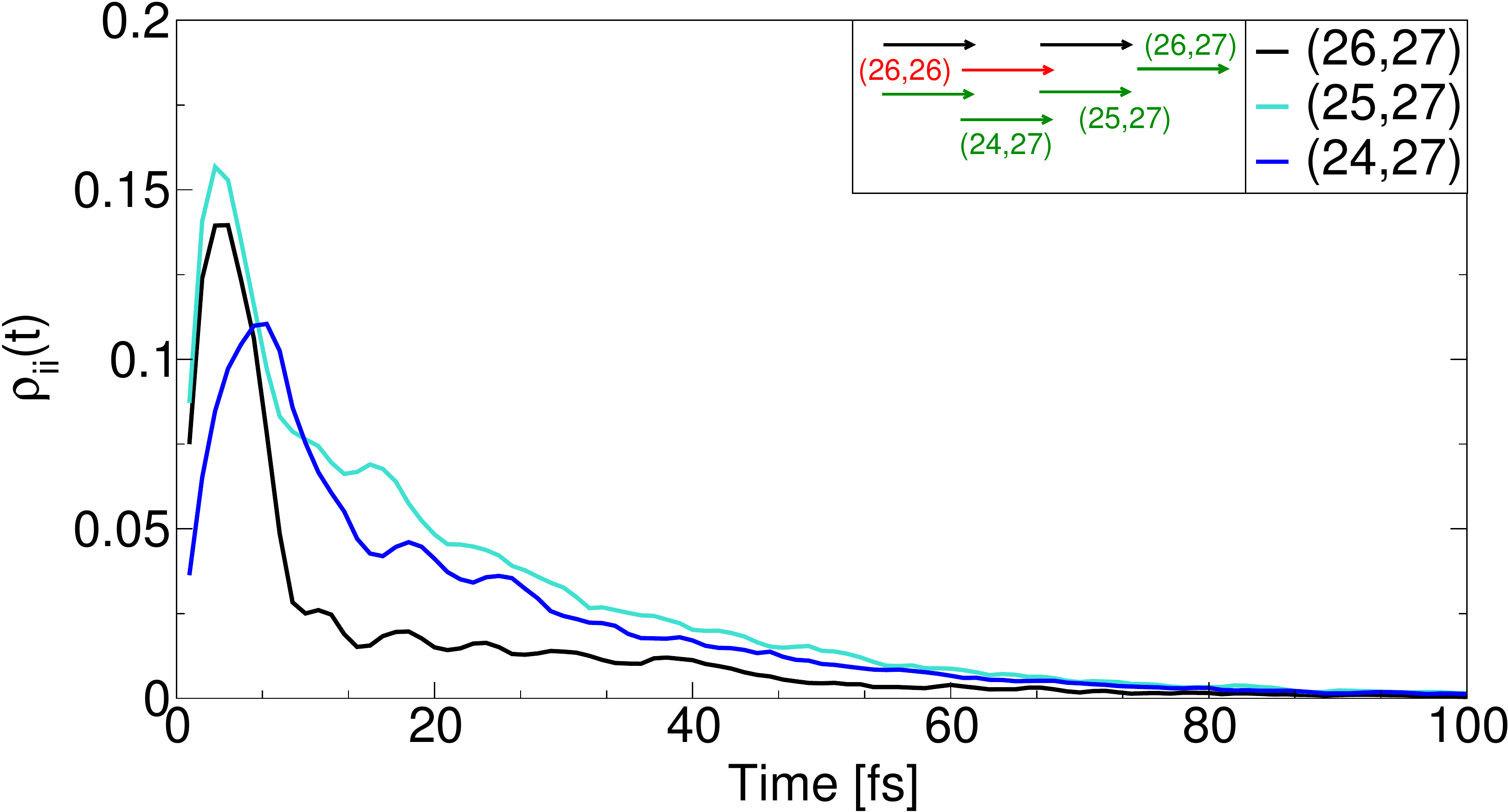}
\par\end{centering}

\caption{Populations of sites neighboring source (26,26) as a function of time
for TDBC with $\sigma=70\,\mathrm{meV}$ and $\Gamma^{\phi}=30\,\mathrm{meV}$.
Population is transferred most rapidly from the central site (26,26)
to site (25,27), the first nearest neighbor and then to site (26,27)
which is in the direction of maximum transport. \label{fig:Single-populations}}
\end{figure}

Based on the discussion in Section \ref{sub:dephasing}, this value
of dynamic disorder represents a lower bound to the exciton dephasing
rate (upper bound to the exciton diffusion length) at room temperature.
One can see that exciton transport is anisotropic and the population
spreads in time following approximately an elliptic shape with major
axis $x$ and minor axis $y$ (such directions are indicated in Fig.
\ref{fig:Brickstone-lattice.-Green-1-1}). The population spreads
about 2 to 3 times faster in the $x$ direction than it does in the
$y$ direction. This can be explained by analizing the direction of
maximum coupling between nearest neighbors. In fact, looking at the
lattice in Fig. \ref{fig:Brickstone-lattice.-Green-1-1} and considering
the molecule placed at the origin of the purple lattice vectors, we
see that it has four first nearest neighbors, one of which is indicated
by the $L_{y}$ vector. The sum of the nearest neighbors coupling
vectors taken in pairs along the $x$ axis is greater than that along
the $y$ axis. The fact that the exciton is transported in the direction
of maximum coupling is also seen in Fig. \ref{fig:Single-populations}
where the population is more rapidly transferred to site (26,27) along
the principle $x$ direction respect to site (24,27) which is along
the $y$ axis. 

By comparing Fig. \ref{fig:Pop1} to Fig. \ref{fig:Cohe1}, we notice
that coherences are spread and suppressed much more rapidly than populations.
However, the principal transport axis remains the same, i.e it is
the major axis of the ellipse. This is true for all examined values
of static and dynamic disorder. At times shorter than about $30\,\mathrm{fs}$
the dynamics is coherent and interesting shapes such as a four leaved
clover at $2\,\mathrm{fs}$ appear in the coherence plot. At longer
timescales incoherent diffusion prevails and these features mostly
disappear. The peaks of these correspond to beats in coherences between
second to fourth nearest neighbors but such beats die off rapidly
over a couple of tens of femtoseconds. While quantitative differences
are observed between the molecules, the general trend is qualitatively
similar to the one shown in the example. As a possible extension of
this initial study of coherent dynamics it would be interesting to
explore the phase directed exciton transport in these two-dimensional
aggregates as has been done for the one dimensional case \cite{Eisfeld2011}. 

While the initial condition (a localized exciton) determines a rather
large population of the central site even at relatively long times,
this does not globally affect the exciton diffusion properties of
the aggregate. Furthermore, the localized excitation can be viewed
as exciton injection from a donor.

\subsection{Diffusion coefficient and diffusion length\label{sub:diff_ris}}

\begin{figure}[h]
\begin{centering}
\includegraphics[angle=-90,width=1\columnwidth]{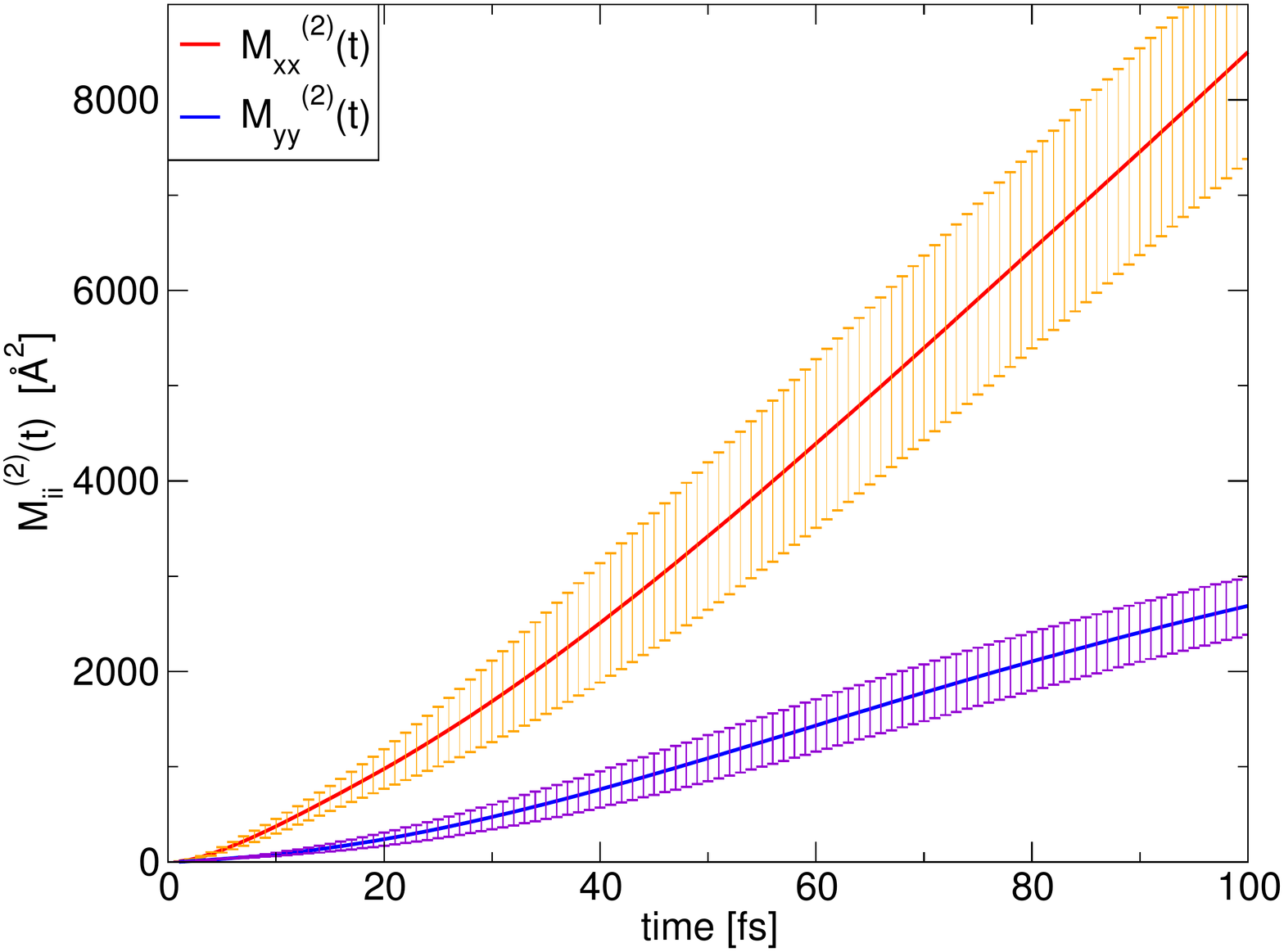}
\par\end{centering}

\caption{Second moment of the exciton wavefunction in time for TDBC aggregate
with static disorder $\sigma=70\,\mathrm{meV}$ and dynamic disorder
$\Gamma^{\phi}=30\,\mathrm{meV}$. Both $xx$ and $yy$ components
are included with their respective error bars. A transition from a
ballistic regime to a diffusive regime is observed at about $30\,\mathrm{fs}$.
For all studied values of static and dynamic disorder these two regimes
are observed and the transition is always at about $20-30\,\mathrm{fs}$.\label{fig:Second-moment-of}}
\end{figure}

The computed wave function provides the most complete source of information
about the exciton dynamics. Its second moment can capture the main
diffusive and ballistic features of the transport. For a homogeneous
system with stochastic dephasing noise andtranslational invariance,
the moments of the wave function in Eq. \ref{eq:full_langevin} can
be written analytically \cite{Reineker1973}. If the exciton is initially
localized on a particular molecule one should expect a ballistic exciton
propagation (the second moment scales quadratically in time, $M^{(2)}\propto t^{2}$)
followed by the diffusive motion (the second moment is linear in time,
$M^{(2)}\propto t$). For systems with static disorder similar transport
regimes should be observed provided that the dynamic noise is strong
enough to overcome exciton localization. To verify this the second
moments of the wave functions each trajectory were computed over an
interval of $100\,\mathrm{fs}$ with a time step of $1\,\mathrm{fs}$
and averaged over a thousand trajectories. In Fig. \ref{fig:Second-moment-of}
an example of $M^{(2)}(t)$ for the TDBC J-aggregate with static disorder
$\sigma=70\,\mathrm{meV}$ and dynamic disorder $\Gamma^{\phi}=30\,\mathrm{meV}$
is shown.
 On timescales shorter than about 30 fs the scaling of $M^{(2)}$is
approximately quadratic, while the longer time dynamics reflects the
diffusive transport. A similar tendency is observed for other molecules
within the whole studied spectrum of $\sigma$ and $\Gamma^{\phi}$
with the free exciton propagation time shrinking down to below 20
fs for large values of $\Gamma^{\phi}$ .

\begin{figure}[h]
\begin{centering}
\includegraphics[angle=-90,width=1\columnwidth]{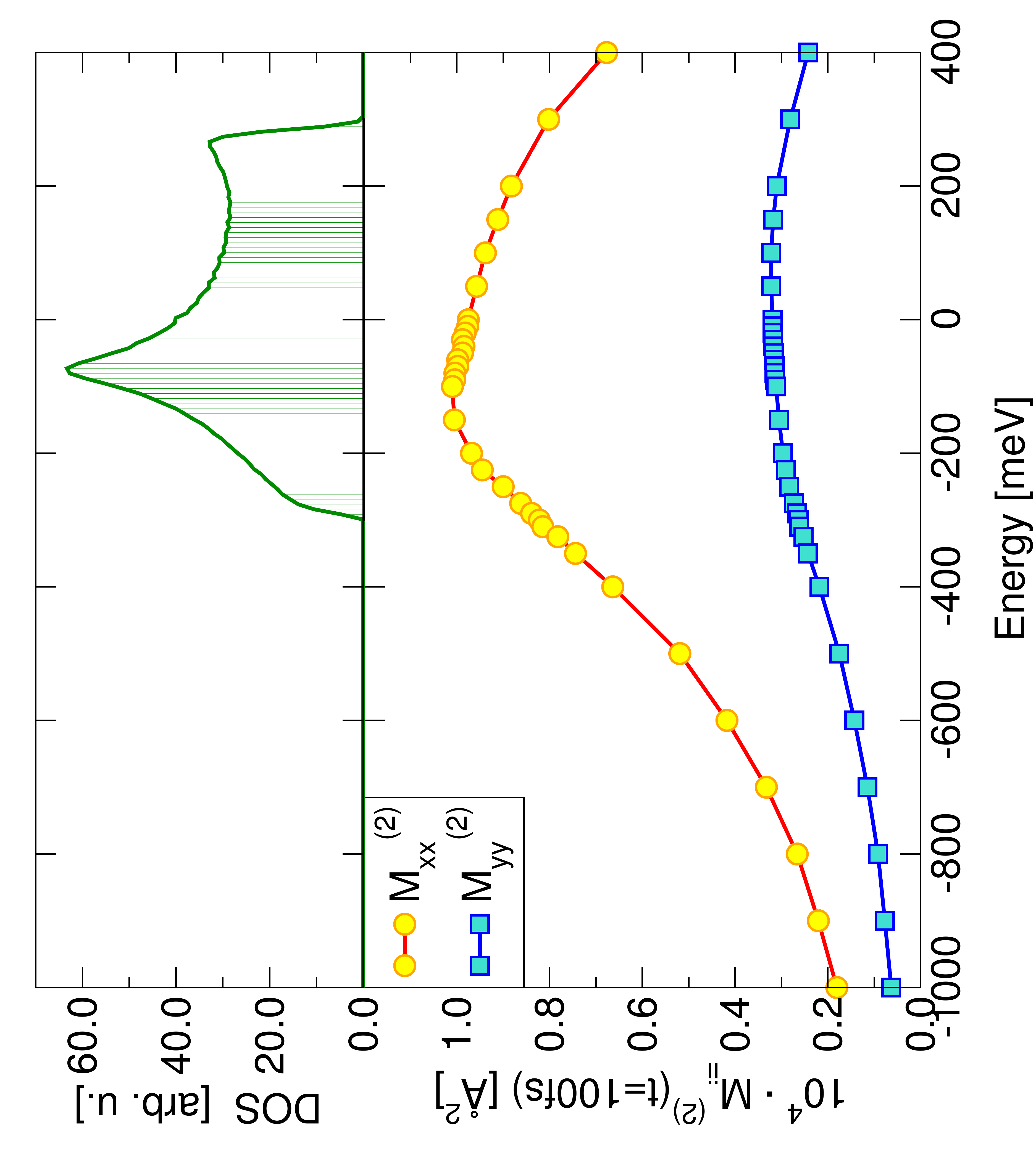}
\par\end{centering}

\caption{Top panel: Density of exciton states as a function of energy of states
computed for an aggregate of 51x51 TDBC molecules and averaged over
1000 realizations of static disorder $\sigma=70\,\mathrm{meV}$. The
zero of energy corresponds to the electronic transition of a single
molecule. The energy of the central site (injection point) in the
lattice is assumed to be zero. Bottom panel: Plot of second moments
of the wave function at time $t=100\,\mathrm{fs}$ as a function of
the initial injection energy for TDBC with static disorder $\sigma=70\,\mathrm{meV}$
and $\Gamma^{\phi}=30\,\mathrm{meV}$.\label{fig:Dependence-on-initial-1}}
\end{figure}

The exciton transport properties depend on the initial injection energy
$E_{in}$. Results showing the second moment as a function of $E_{in}$
together with the exciton density of states (DOS) are reported in
Fig. \ref{fig:Dependence-on-initial-1}. 
 We see that there is a maximum in the moments somewhere between the
J-band ($E_{in}=-281\,\mathrm{meV}$) and the monomer band ($E_{in}=0\,\mathrm{meV}$).
This is true at all times and the position of the maximum is different
for the $xx$ component than it is for the $yy$ component. This can
be correlated to the large density of states in that energy interval
(Fig. \ref{fig:Dependence-on-initial-1} top panel). The high density
of states is not sufficient to explain the exact position of the maximum,
in fact, while the density of states has a maximum at about $-72\:\mathrm{meV}$,
the second moment is peaked at about $-170\,\mathrm{meV}$ for the
$xx$ component. The same trend in the second moments is observed
for TC, and the only apparent difference is that the values of the
second moment are smaller. One aspect to keep into consideration is
that although there is a large density of states around $60-80\,\mathrm{meV}$,
these states mostly have small oscillator strengths (See Fig. \ref{fig:Calculated-absorption-spectrum}).
To investigate further the origin of the differences in these maxima
one should look into the structure and spatial overlap of the exciton
states.

\begin{figure}[h]
\begin{centering}
\includegraphics[angle=-90,width=1\columnwidth]{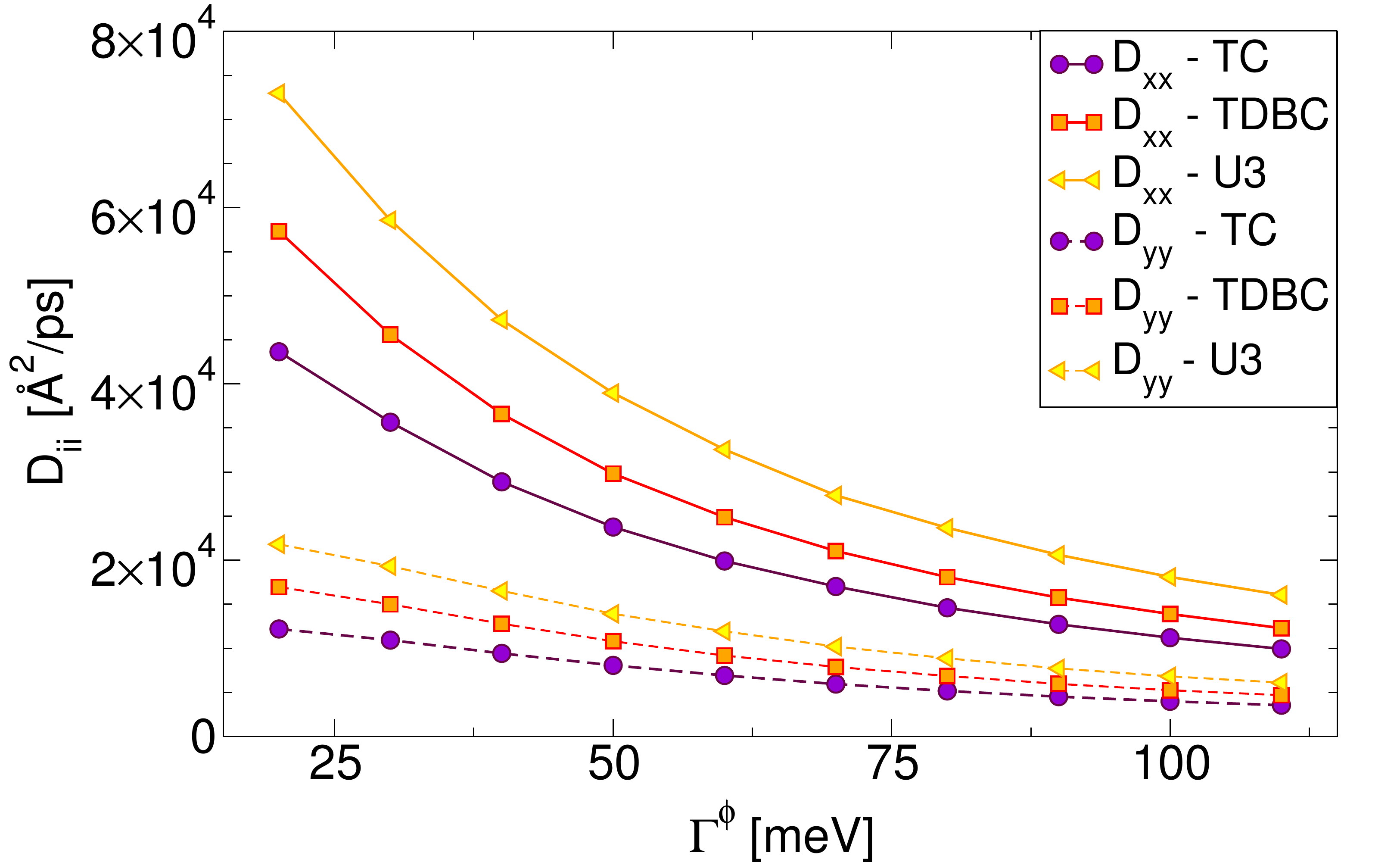}
\par\end{centering}

\caption{Comparison of diffusion constants as a function of dynamic disorder
$\Gamma^{\phi}$ for TC, TDBC and U3 with static disorder $\sigma=70\,\mathrm{meV}$.
As dynamic disorder is increased, the diffusion constants decrease.
Further, $xx$ diffusion constants are larger than $yy$ components
for all values of dynamic disorder. Finally, diffusion is larger for
U3 than it is for TDBC and TC. This can be explained by the molecular
coupling parameters as described in the text. \label{fig:Comparison-of-diffusion}}
\end{figure}

To estimate the exciton diffusion coefficients the linear fit was
taken over the time interval $\Delta t_{1}=30-100$ fs to exclude
the initial ballistic propagation of the exciton. For most values
of static and dynamic disorder, the boundary effects associated with
the finite size of the simulated lattice is negligible on the timescale
of 100 fs. However, for weak dynamic noise the exciton wavefunction
reaches the boundary of the simulated lattice along the y-axis at
about 70 fs. In such cases the $yy$ component of the diffusion coefficient
$D_{yy}$ was fitted on the time range $30-70$ fs. With the initial
condition of injection into the J-band, the diffusion coefficients
were calculated using 

\begin{equation}
M_{ii}^{(2)}(t)=2D_{ii}t
\end{equation}
for each type of aggregate. The results are shown in Fig. \ref{fig:Comparison-of-diffusion}.
Two distinct characteristics emerge from this plot. First of all,
we notice that diffusion is greater for U3 than it is for TDBC which
in turn is greater than that of the TC aggregate. This can be explained
by looking at the physical characteristics of the molecules (Tab.
\ref{tab:dye-properties}). U3 has the largest transition dipole,
this leads to stronger coupling between the monomers and thus to more
rapid exciton density transfer. The diffusion coefficients normalized
with respect to the square of the corresponding transition dipole
(not shown in the figures) are similar for two molecules, TDBC and
U3, while the normalized diffusion for TC is still higher. We attribute
this difference to the closer packing of TC molecules. 
 This trend is robust to both static and dynamic disorder, so long
as both are finite and not so large as to lead to localization. This
implies that independently of temperature, within the studied interval,
the transport efficiency is dictated by the specific physical properties
of the molecules. Of course, the geometric arrangement is a criteria
which can alter this trend since it will explicitly modify the couplings.
Then secondly, as noticed from the wavefunction propagation, diffusion
is faster along the $x$ axis than it is along the $y$ axis. This
difference is largest for small values of disorder (about a 3-fold
difference)) where the propagation is fastest, in the quasi ballistic
regime. Going to larger disorder, the wavefunction spreads much slower
and propagation is reduced in both directions. 

\begin{figure*}
\begin{centering}
\textbf{a}\includegraphics[width=1\columnwidth]{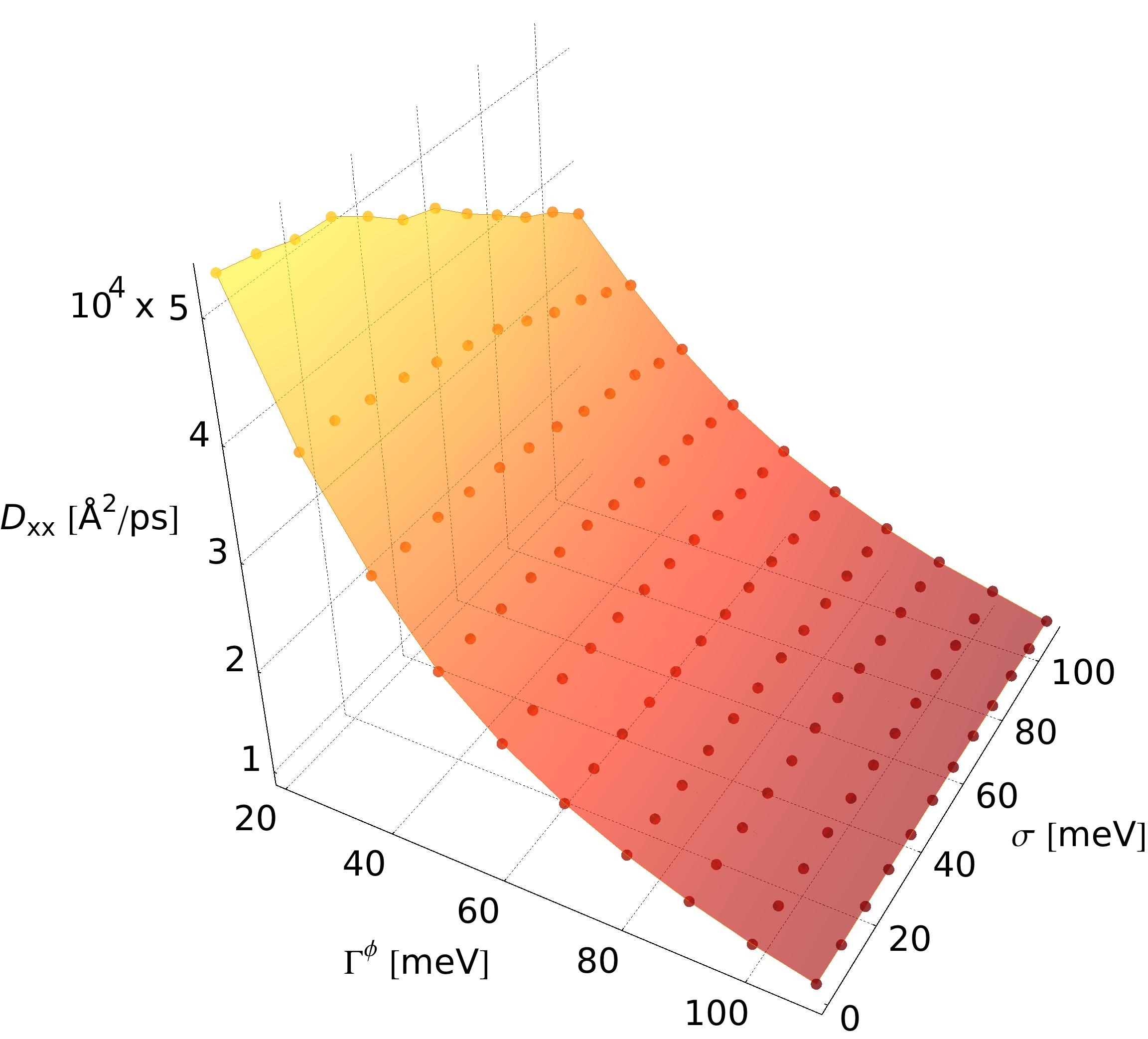}\includegraphics[width=1\columnwidth]{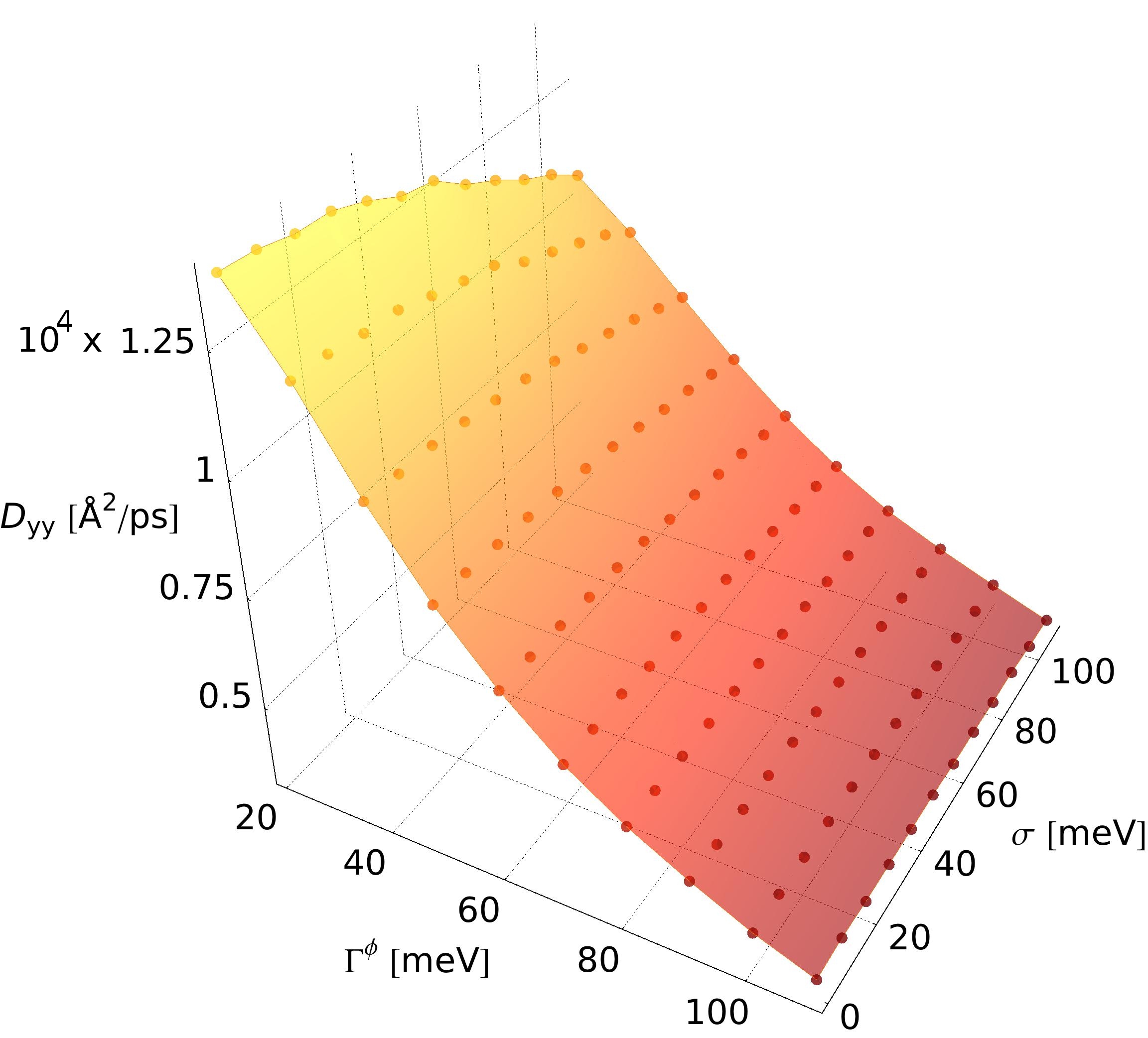}
\par\end{centering}

\centering{}\textbf{b}\includegraphics[width=1\columnwidth]{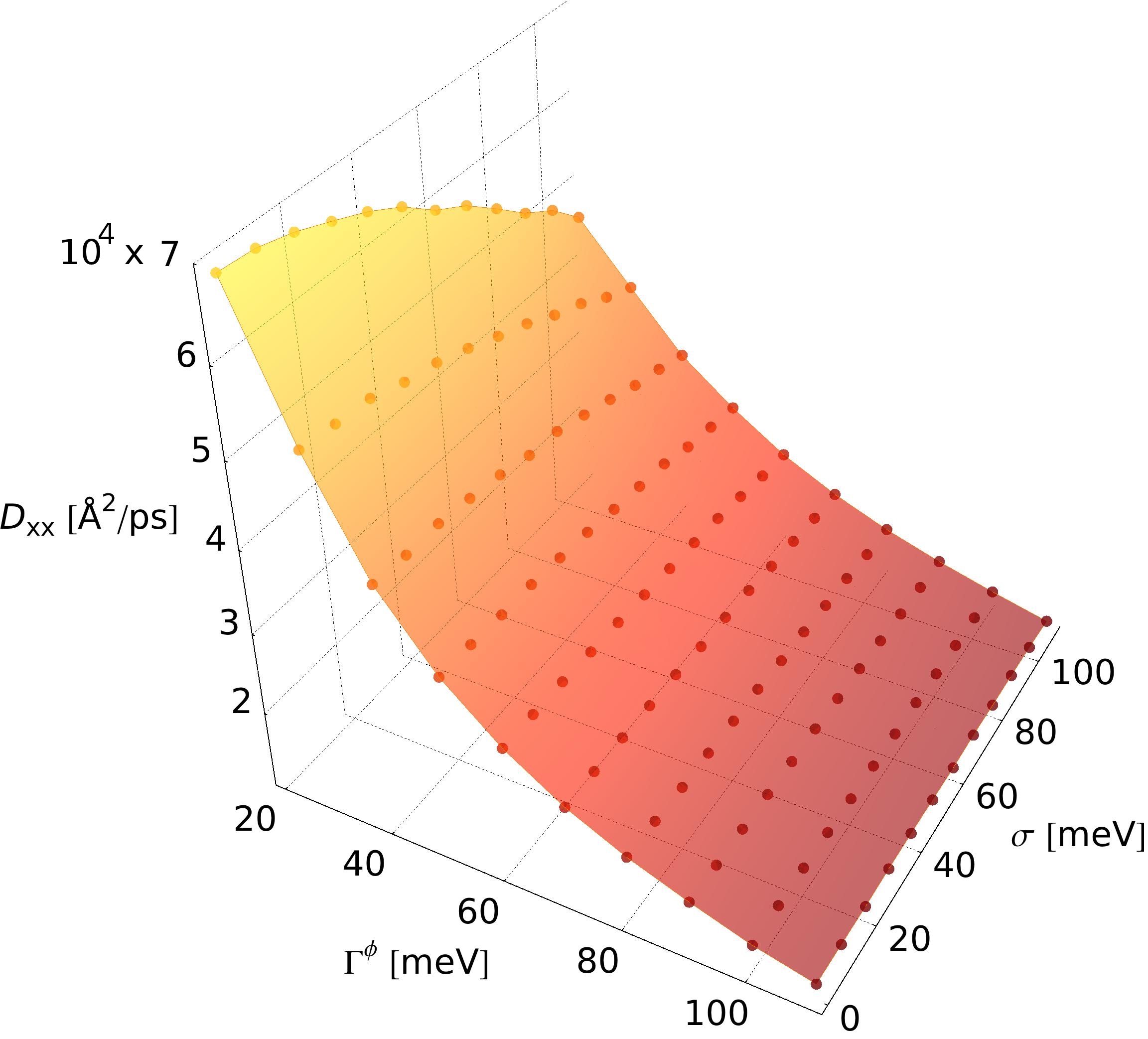}\includegraphics[width=1\columnwidth]{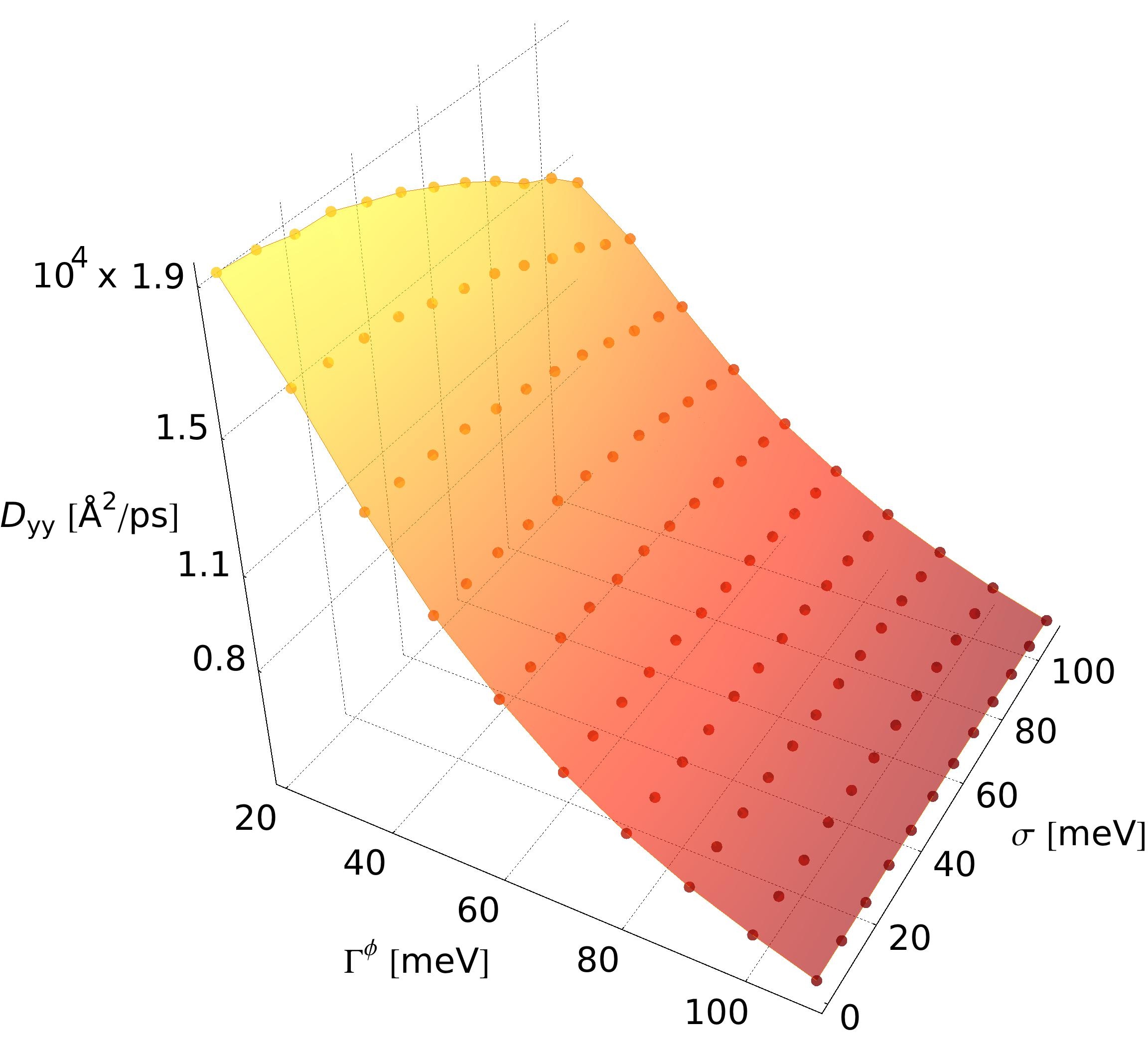}\caption{Diffusion coefficients of singlet excitons in 2D J-aggregates of TC
(panel a) and TDBC (panel b) as functions of static $\sigma$ and
dynamic disorders $\Gamma^{\phi}$. The initial condition for propagation
was injection at $E_{in}=-281\,\mathrm{meV}$ for TDBC and $E_{in}=-287\,\mathrm{meV}$
for TC. The red color corresponds to smaller diffusion while the yellow
indicates larger diffusion coefficients.\label{fig:Diffusion_3D_maps}}
\end{figure*}

Finally, we investigated the transport as a function of static disorder
as well. The three dimensional surface plots for TC and TDBC are shown
in Fig. \ref{fig:Diffusion_3D_maps}. We notice that diffusion is
strongly dependent on $\Gamma^{\phi}$ and much less dependent on
$\sigma$. For each fixed value of $\Gamma^{\phi}$, the largest variation
of $D_{ii}$ over the $\sigma$ interval is of about $30\%$ of the
largest values. On the other hand, the largest variation of $D_{ii}$
at fixed $\sigma$ and varying $\Gamma^{\phi}$ is of the order of
$80\%$ of the largest value. Due to the presence of static disorder
the dependence of the diffusion coefficient on the dephasing rate
$\Gamma^{\phi}$ deviates from the conventional $D\propto1/\Gamma^{\phi}$
derived within the Haken-Strobl-Reineker model for homogeneous systems
\cite{Reineker1973}. 

Recently, Akselrod et. al. \cite{Akselrod2010}, have estimated the
exciton diffusion length in thin-film J-aggregates of TDBC molecules
based on an exciton-exciton annihilation experiment at room temperature.
In that study, the exciton lifetime obtained from time-dependent photoluminescence
was determined to be $\tau_{\mathrm{exp}}=45\,\mathrm{ps}$, and an
expression for three-dimensional exciton diffusion was used to determine
the annihilation rate. However, the sulphonated group side chains
are about $6\,\textrm{\AA}$ long and in addition, in the growth process
a layer of polymer molecules is introduced between the J-aggregate
layers. As a result the spacing between the monolayers of fluorescent
dyes is several times larger than the distance between nearest neighbor
molecules in a layer. Therefore we can assume that diffusion in these
aggregates is two dimensional rather than three dimensional. Using
a 2D model, we find that the experimental exciton diffusion length
is about $\mathcal{\ell}_{\mathrm{exp}}\approx60$ nm. We can estimate
the exciton diffusion length along the $i$-th direction using the
measured lifetime and a computed diffusion coefficient as 

\begin{equation}
\mathcal{\ell}_{i}=\sqrt{2D_{ii}\tau_{\mathrm{exp}}}.
\end{equation}

For the value of the dynamic disorder $\Gamma^{\phi}=$30 meV, which
should approximately correspond to room temperature, as discussed
in Section \ref{sub:dephasing}, and the static disorder $\sigma=70\,\mathrm{meV}$
we find $\mathcal{\ell}_{x}\approx200$ nm and $\mathcal{\ell}_{y}\approx100$
nm. These values are in a good qualitative agreement with the measured
one. However, the quantitative discrepancy can be due to a number
of factors such as different lattice constants of the aggregate, additional
exciton dephasing and relaxation channels, and also domain boundaries
in the experimental structures. All of these aspects can be incorporated
into the model provided that one can extract the actual parameters
should from experiments.

\section{Conclusions\label{sec:conclusions}}

In this article, a mixed model combining an open quantum systems approach
to \emph{ab-initio} calculations has been employed to gain insight
on the exciton dynamics of thin-film J-aggregates. This model can
capture both coherent and incoherent transport and allows for a detailed
study of transport parameters such as diffusion coefficients and diffusion
length. The role of the initial state of the system on transport can
also be captured. Further one can investigate all these aspects as
a function of the structure of the aggregates through the lattice
parameters. 

As an example the model was applied to three different cyanine dye
aggregates. Within this model we conclude that transport depends explicitly
on the molecular properties of the monomers which compose the aggregate.
In particular, for molecules packed in a brickstone arrangement, transport
increases with the monomers' transition dipoles and hence with the
coupling between monomers. Furthermore, the coupling induces a preferential
direction for transport which leads to an anisotropic spread of populations
and coherences. Such directionality is robust to both static and dynamic
disorder within the investigated ranges and does not change for different
molecules. This model has permitted the identification of timescales
for the different transport regimes. A ballistic regime is present
for all values of disorder at times smaller than $20\,\mathrm{fs}$
while afterwards a diffusive regime is observed. The transport is
also determined by the choice of the initial condition, at a specific
value of injection energy a maximum in diffusion is observed. Investigation
on the origin of the exact position of this maximum are planned but
qualitatively it can be explained by the large DOS located between
J-band and monomer transition. 

The obtained diffusion length is in good agreement with experimental
results, however a more accurate comparison would be possible only
including relaxation in the excited state manifold. The model does
not account for exciton domains which would reduce diffusion, and
further we cannot investigate the low temperature regime. Work in
these directions is currently in progress in our research group. 

The efficient exciton transport observed in these thin-film aggregates
and the possibility of tuning this transport by the choice of monomers
or by selecting the initial condition makes these aggregates good
candidates for devices where large exciton diffusion lengths are sought.
In particular coupling them to optical micro-cavities \cite{Saikin2011}
opens the road to a range of control possibilities which we plan to
study in future work. Coupling these large-exciton diffusion length
materials to high-hole mobility materials \cite{Sule2011} might also
provide some advantages for technological applications such as all-organic
photon detectors.
\begin{acknowledgments}
We appreciate useful discussions with G. Akselrod, D. Arias, R. Olivares-Amaya
and D. Rappoport. Further the authors would especially like to thank
A. Eisfeld for the many fruitful comments which greatly enriched this
project. 

This work was supported by the Defense Threat Reduction Agency under
Contract No HDTRA1-10-1-0046. S. V. acknowledges support from the
Center for Excitonics, an Energy Frontier Research Center funded by
the U.S. Department of Energy, Office of Science and Office of Basic
Energy Sciences under Award Number DE-SC0001088 as well as support
from the Defense Advanced Research Projects Agency under award number
N66001-10-1-4060. M-H. Yung also acknowledges the Croucher Foundation
for support. Further, A. A.-G. is grateful for the support of the
Camille and Henry Dreyfus Foundation and the Alfred P. Sloan Foundation. 
\end{acknowledgments}
\begin{appendix}

\section{Single molecule Hamiltonian \label{sec: app_1}}

\begin{figure}
\centering{}\includegraphics[width=0.7\columnwidth]{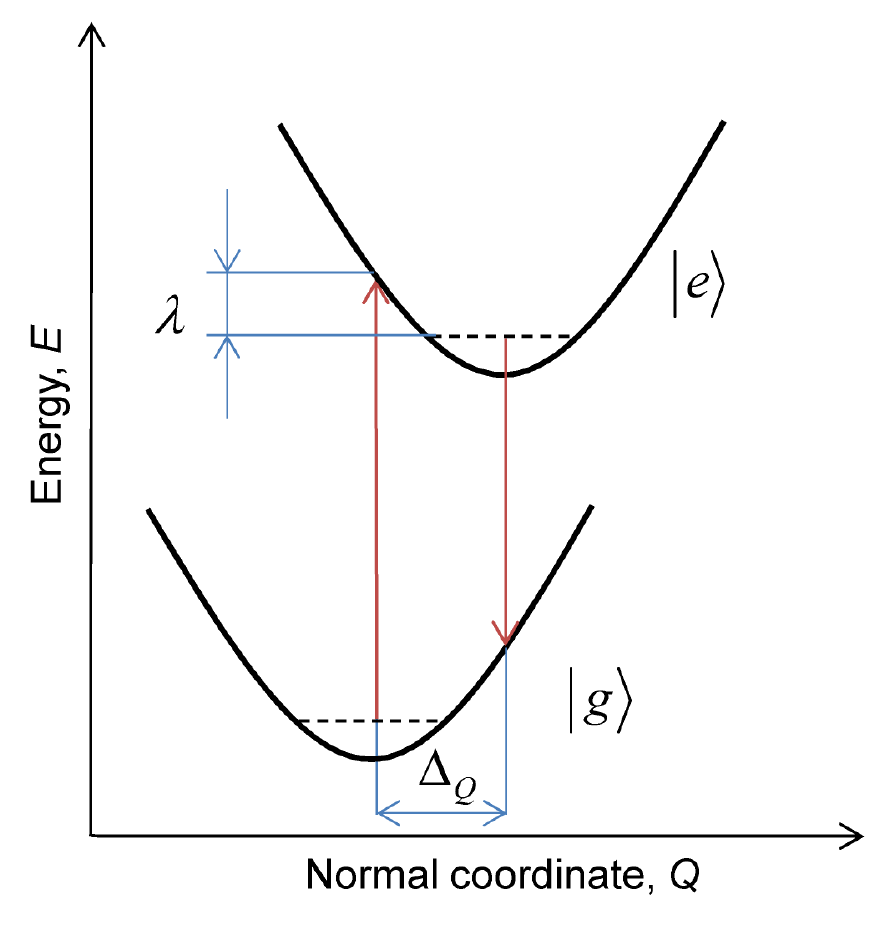}\caption{Schematic energy profile of a single vibrational mode in a two-level
molecule. $\lambda$ is a reorganization energy and $\Delta_{Q}$is
a normal coordinate displacement.\label{fig:two-level}}
\end{figure}

In the harmonic approximation, the single-molecule electron excitation
Hamiltonian can be written as 
\begin{equation}
\hat{H}_{{\rm mol}}=\sum_{i}\left(\Omega_{i}+\sum_{q}\omega_{iq}\hat{b}_{iq}^{\dagger}\hat{b}_{iq}\right)|i\rangle\langle i|,\label{eq:Monomer Hamiltonian-1-1}
\end{equation}
where $\Omega_{i}$ is the energy of $i$-th electronic transition,
$b_{iq}^{\dagger}$ and $b_{iq}$ are creation and annihilation operators
for the $q$-th vibration and $\omega_{iq}$ is the frequency of the
corresponding vibrational mode. Within the visible range of the spectrum,
the lowest electronic transition in fluorescent dyes has the largest
oscillator strength. This transition determines the formation of a
J-band in molecular aggregates. An approximation involving only the
two lowest electronic states - the ground $|g\rangle$ and the first
excited $|e\rangle$ states of a monomer - can provide good description
of the excitation dynamics as long asthe second excited state is far
enough from the first state. This approximation has been used previously
to model excitons in J-aggregates \cite{Knoester1993,Lemaistre2008a,Lemaistre2007}.
In the two-level approximation Eq. \ref{eq:Monomer Hamiltonian-1-1}
simplifies to

\begin{equation}
\hat{H}_{{\rm mol}}=\tilde{\Omega}_{e}|e\rangle\langle e|+\sum_{q}\omega_{q}\hat{b}_{q}^{\dagger}\hat{b}_{q}+\sum_{q}\kappa_{q}|e\rangle\langle e|\left(\hat{b}_{q}^{\dagger}+\hat{b}_{q}\right)\label{eq:Monomer two-level-1-1}
\end{equation}
where we have assumed that the frequencies of the vibrational modes
are the same for all electronic states, and the renormalized electronic
transition frequency $\tilde{\Omega}_{e}=\Omega_{e}+\sum_{q}\lambda_{q}$
involves the reorganization energies $\lambda_{q}$ associated with
each mode $q$. The exciton-vibration coupling coefficient is defined
as $\kappa_{q}=\lambda_{q}/\Delta_{q}$, where the boson operators
associated with different electronic states are related by $\hat{b}_{eq}=\hat{b}_{gq}+\hat{\Delta}_{q}$.
In Eq. \ref{eq:Monomer two-level-1-1} all vibrational modes correspond
to the electronic~ground state and the corresponding ground state
indices are omitted for simplicity. The phonon creation/annihilation
operator displacement $\hat{\Delta}_{q}$ is related to the normal
coordinate displacement operator $\hat{\Delta}_{Qq}$ as $\hat{\Delta}_{q}=\hat{\Delta}_{Qq}\cdot\sqrt{2\omega_{q}/\hbar}$.
A schematic diagram of a two-level molecule is shown in Fig. \ref{fig:two-level}. 

\section{Exciton manifolds \label{sec: app_2}}

\begin{figure}
\centering{}\includegraphics[width=0.6\columnwidth]{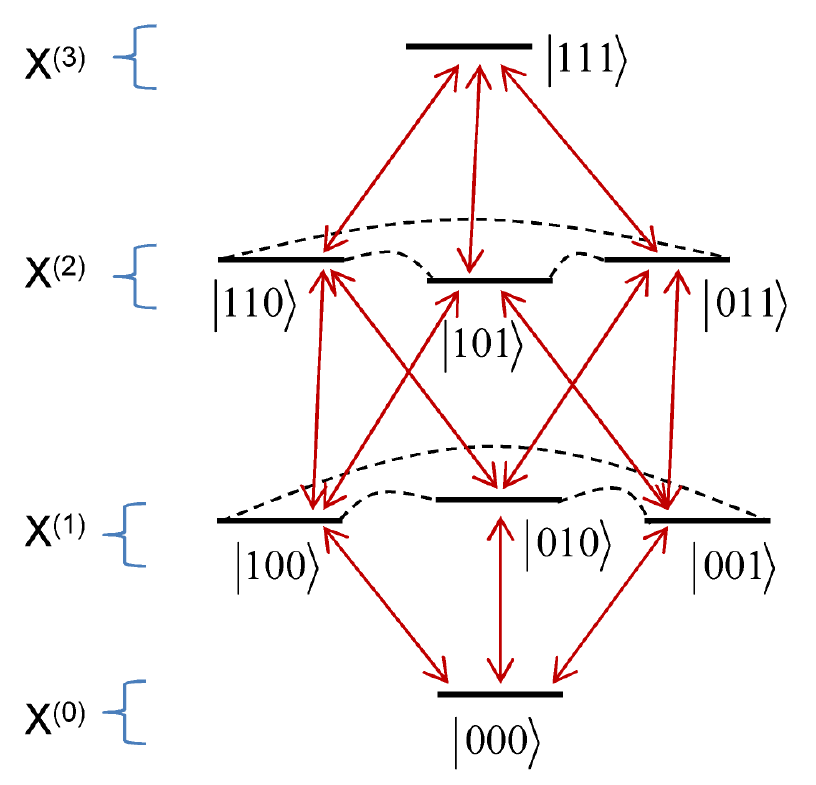}\caption{Schematic diagram of exciton manifolds - ground state $X^{(0)}$,
single exciton $X^{(1)}$, bi-exciton $X^{(2)}$, and tri-exciton
$X^{(3)}$ - in an aggregate of threetwo-level molecules. The arrows
between the manifolds show allowed exciton creation and relaxation
transitions, and the dashed lines show Coulomb couplings between the
states within a manifold. The exciton-exciton annihilation processes
are shown.\label{fig:Schematic-diagram-1-1} }
\end{figure}

For an aggregate consisting of $N$ molecules all $2^{N}-1$ excited
states can be classified into $N$ different exciton manifolds - single
exciton $X^{(1)}$, bi-exciton $X^{(2)}$, etc, see Fig. \ref{fig:Schematic-diagram-1-1}.
The energy separation between the manifolds is of the same order of
magnitude as that of the single exciton energy. The exiton manifolds
are coupled by exciton relaxation or annihilation processes and also
by interaction with external radiation. Within each subspace, the
exciton states of monomers are coupled to each other by the Coulomb
interaction, which gives rise to a resonant excitation transfer between
molecules \cite{Foersterbook}. The same Coulomb interaction also
renormalizes the energies of single monomer excitations. However,
transitions between different exciton manifolds induced by the Coulomb
interaction can be ignored. Thus, one can study single exciton dynamics
by constraining the exciton space to the single exciton manifold $X^{(1)}$
and the ground state $X^{(0)}$. In Fig. \ref{fig:Schematic-diagram-1-1}
we show a schematic diagram of exciton manifolds together with the
allowed transitions for a system of three coupled molecules. Unlike
molecular crystals, delocalized exciton bands are not formed in molecular
aggregates due to a strong competition between the Coulomb interaction,
which tends to delocalize the exciton and to the exciton-vibration
coupling and structure disorder, which favor localization.

\end{appendix}

%

\end{document}